\documentclass[onecolumn,romanappendices]{IEEEtran}

\usepackage{etex}
\usepackage[cmex10]{amsmath}
\usepackage{amssymb}
\usepackage{amsthm}
\usepackage[hidelinks]{hyperref}
\usepackage[capitalize,noabbrev]{cleveref}
\usepackage{cite}
\usepackage{url}
\usepackage{enumerate}
\usepackage[T1]{fontenc}
\usepackage{booktabs}
\usepackage{multirow}
\usepackage[font=footnotesize]{caption}
\usepackage[font=footnotesize]{subcaption}
\usepackage{fullpage}
\usepackage{graphicx}
\usepackage[ruled,vlined]{algorithm2e}
\usepackage{tikz}
\usepackage{tikz-qtree}
\usetikzlibrary{backgrounds,fit}
\usetikzlibrary{matrix,arrows,shapes,positioning,chains,scopes,decorations.markings,calc,patterns}

\usepackage{pgfplots}
\usepackage{pgfplotstable}
\usepackage{booktabs}

\pgfdeclarelayer{background}
\pgfsetlayers{background,main}

\usepackage[colorinlistoftodos,textsize=small]{todonotes}

\newcommand{\ie}{{i.e.}\xspace}
\newcommand{\ZZ}{{\mathbb{Z}}\xspace}
\newcommand{\RR}{{\mathbb{R}}\xspace}
\newcommand{\NN}{{\mathbb{N}}\xspace}
\newcommand{\Bset}{{\mathcal{B}}\xspace}
\newcommand{\Cset}{{\mathcal{C}}\xspace}

\newcommand{\Qset}{{\mathcal{Q}}\xspace}
\newcommand{\Rset}{{\mathcal{R}}\xspace}
\newcommand{\Sset}{{\mathcal{S}}\xspace}
\newcommand{\suchthat}{{\it s.t.}\xspace}
\newcommand{\size}{{\mathsf{Size}}\xspace}
\newcommand{\area}{{\mathsf{Area}}\xspace}
\newcommand{\location}{{\mathsf{Loc}}\xspace}
\newcommand{\ellmax}{\ell^\text{max}}
\newcommand{\region}{\mathsf{Reg}}
\newcommand{\QuotientContainers}{\mathsf{QuoCon}}
\newcommand{\RemainderRegions}{\mathsf{RemReg}}

\makeatletter
\newcommand{\removelatexerror}{\let\@latex@error\@gobble}
\makeatother

\newtheorem{claim}{Claim}
\newtheorem{lemma}{Lemma}
\newtheorem{theorem}{Theorem}

\theoremstyle{definition}\newtheorem{definition}{Definition}
\theoremstyle{remark}\newtheorem{example}{Example}

\DeclareMathOperator*{\argmin}{arg\,min}

\DeclareMathOperator{\im}{Im}

\begin{document}

\title{Decision Procedure for the Existence of Two-Channel Prefix-Free Codes}

\author{
	\IEEEauthorblockN{Hoover~H.~F.~Yin, Ka~Hei~Ng, Yu~Ting~Shing, Russell~W.~F.~Lai, and Xishi Wang}
	\thanks{H.~Yin and X.~Wang are with the Department of Information Engineering, The Chinese University of Hong Kong, Hong Kong. Email: \{yhf015, wx116\}@ie.cuhk.edu.hk}
	\thanks{K.~Ng is with the Department of Physics, The Chinese University of Hong Kong, Hong Kong. Email: kaheicanaan@gmail.com}
	\thanks{Y.~Shing is with the Mathematics Panel, St. Francis Xavier's College, Hong Kong. Email: js@sfxc.edu.hk}
	\thanks{R.~Lai is with the Chair of Applied Cryptography, Friedrich-Alexander-Universit\"at Erlangen-N\"urnberg, Germany. Email: russell.lai@cs.fau.de}
	\thanks{Part of the work of H.~Yin was done when he was with the Institute of Network Coding, The Chinese University of Hong Kong, Hong Kong, which was supported by a grant from the University Grants Committee of the Hong Kong Special Administrative Region (Project No. AoE/E-02/08).}
	\thanks{The work of R.~Lai was supported by the German research foundation (DFG) through the collaborative research center 1223, and by the state of Bavaria at the Nuremberg Campus of Technology (NCT).
		NCT is a research cooperation between the Friedrich-Alexander-Universit\"at Erlangen-N\"urnberg (FAU) and the Technische Hochschule N\"urnberg Georg Simon Ohm (THN).}
}

\maketitle

\begin{abstract}
	The Kraft inequality gives a necessary and sufficient condition for the existence of a single channel prefix-free code.
	However, the multichannel Kraft inequality does not imply the existence of a multichannel prefix-free code in general.
	It is natural to ask whatever there exists an efficient decision procedure for the existence of multichannel prefix-free codes.
	In this paper, we tackle the two-channel case of the above problem by relating it to a constrained rectangle packing problem.
	Although a general rectangle packing problem is NP-complete, the extra imposed constraints allow us to propose an algorithm which can solve the problem efficiently.
\end{abstract}

\thispagestyle{empty}

\section{Introduction}

\emph{Prefix-free codes}, also known as \emph{prefix codes}, are a class of uniquely decodable codes which has the property that a codeword can be decoded without referring to the symbols of any future codewords.
In other words, a codeword of a prefix code can be recognized instantaneously once all the symbols of that codeword are received.
There exist prefix codes which are optimal codes, e.g., the Huffman codes \cite{huffman}.

There are different constructions of single channel prefix codes for data compression, e.g., Shannon coding \cite{shannon}, Shannon-Fano coding \cite{fano}, and Huffman coding \cite{huffman}.
Among these examples, Huffman coding always produce an optimal code \cite{huffman}.
We can construct a Huffman code in linear time when the statistical information of the data to be compressed is sorted \cite{linear_huffman}.
Other than data compression, prefix codes are also used in applications like country calling codes \cite{itu} and UTF-8 encoding \cite{utf8}.

It is well-known that the Kraft inequality \cite{kraft} 
gives a necessary condition for a single channel source code to be uniquely decodable~\cite{mcmillan}.
This result was generalized into a multichannel case where the channels use a \emph{homogeneous alphabet size}, \ie, all channels use the same alphabet size, in \cite{yao10}.
The case for \emph{heterogeneous alphabet sizes}, \ie, the channels can use different alphabet sizes, was not investigated in \cite{yao10}.
On the other hand, the Kraft inequality gives a sufficient condition for the existence of a single channel prefix code~\cite{info_book}.
For the multichannel case, however, the sufficient condition does not hold in general even for homogeneous alphabet size~\cite{yao10}.
A natural problem thus arises: Does there exist an efficient decision procedure for the existence of multichannel prefix codes when the channels use heterogeneous alphabet sizes?

The decision procedure aims to close the gap where the multichannel Kraft inequality fails.
Precisely, the decision procedure takes a finite multiset of finite codeword lengths as an input, and decides the existence of a prefix code where the multiset of codeword lengths of the prefix code equals to the input multiset exactly.

In this paper, we generalize the multichannel Kraft inequality for channels using heterogeneous alphabet sizes and illustrate the failure of the sufficient condition geometrically.
We also present the relation between the existence of a prefix code and the existence of solutions of a constrained rectangle packing problem.
We then tackle the two-channel case of the problem of deciding the existence of prefix codes via a reduction to a constrained rectangle packing problem.
Although a general rectangle packing problem is NP-complete~\cite{rectangle}, the constrained version we are interested in can be solved efficiently by our proposed algorithm.

\section{Multichannel Source Codes and Multichannel Prefix-Free Codes}

We describe briefly the multichannel source codes and multichannel prefix-free codes proposed in \cite{yao10} and give some minor generalizations for heterogeneous alphabet sizes.

Let there be $n$ channels.
Denote the alphabet of the information source $Z$ by $\mathcal{Z}$.
The alphabet used in the $i$-th channel is denoted by $\mathcal{Z}_i$ where $i = 1, 2, \ldots, n$.
Let $q_i = |\mathcal{Z}_i|$ be the size of the alphabet.\footnote{The work in \cite{yao10} only considered $q_1 = q_2 = \ldots = q_n$.}
Define $\mathcal{Z}_i^0 = \{\epsilon\}$ and $\mathcal{Z}_i^j = \{wv : w \in \mathcal{Z}_i^{j-1}, v \in \mathcal{Z}_i \}$ for $j \ge 1$, where $\epsilon$ is the empty string.
In other words, $\mathcal{Z}_i^j$ is the set of all possible strings of length $j$ formed by the alphabets in $\mathcal{Z}_i$.
The collection of all concatenations of alphabets from $\mathcal{Z}_i$ is the set $\mathcal{Z}_i^\ast = \bigcup_{j = 0}^\infty \mathcal{Z}_i^j$, 
\ie, $\mathcal{Z}_i^\ast$ is the set containing all possible finite word sequences formed by the alphabets in $\mathcal{Z}_i$, including the empty word.

\begin{definition}[Multichannel Source Codes]
An $n$-channel \emph{source code} $\mathcal{Q}$ for the source random variable $Z$ is a mapping from $\mathcal{Z}$ to $\prod_{i = 1}^n \mathcal{Z}_i^\ast$.
Every element in $\prod_{i = 1}^n \mathcal{Z}_i^\ast$ is called a \emph{word}.
For any \emph{source symbol} $z \in \mathcal{Z}$, $\mathcal{Q}(z)$ is the \emph{codeword} for $z$.
The image $\im(\mathcal{Q}) \subseteq \prod_{i = 1}^n \mathcal{Z}_i^\ast$ of $\mathcal{Q}$ is called the \emph{codebook}.
\end{definition}

\begin{definition}[Multichannel Prefix-Free Codes] \label{def:prefix}
	Two codewords are \emph{prefix-free} to each other if and only if there exists at least one channel $i$ such that the $i$-th component of the two codewords are prefix-free to each other.
	An $n$-channel \emph{prefix-free code} $\mathcal{Q}$ is an $n$-channel source code where $\im(\mathcal{Q}) \subseteq \prod_{i = 1}^n \mathcal{Z}_i^\ast$ such that every pair of codewords in $\im(\mathcal{Q})$ are prefix free to each other.
\end{definition}

When a codeword is transmitted, the $i$-th component of the codeword is transmitted through the $i$-th channel.
When more than one codewords are transmitted, the codewords are concatenated channel-wise.
The boundaries of the codewords are thus not explicit anymore.
In order to distinguish the boundaries, we are interested in a class of source codes called the \emph{uniquely decodable code}.

\begin{definition}[Uniquely Decodable Codes]
	For any two distinct finite sequences of source symbols, if their finite sequences of codewords are different, then the source code is a \emph{uniquely decodable code}.
\end{definition}

Similar to the single channel Kraft inequality, the multichannel Kraft inequality gives a necessary condition for a source code to be uniquely decodable.
Suppose there are $m$ codewords in a $n$-channel source code $\mathcal{Q}$.
Let $\ell_i^j$ be the length of the $j$-th codeword in the $i$-th channel, where $j = 1, 2, \ldots, m$.
The \emph{codeword length} of the $j$-th codeword is defined as a tuple $(\ell_1^j, \ldots, \ell_n^j)$.

\begin{theorem}[Kraft Inequality] \label{thm:kraft}
	If $\mathcal{Q}$ is uniquely decodable, then the lengths of its codewords satisfy
\begin{equation} \label{eq:kraft}
	\sum_{j = 1}^m \prod_{i = 1}^n q_i^{-\ell_i^j} \le 1.
\end{equation}
\end{theorem}

\begin{IEEEproof}
	We can use a similar technique proposed in \cite{mcmillan_proof} to prove the generalized Kraft inequality.
	See Appendix~\ref{sec:kraft_entropy}.
\end{IEEEproof}

When we use a homogeneous alphabet sizes, the multichannel Kraft inequality becomes the version $$\sum_{j = 1}^m q_1^{\sum_{i = 1}^n -\ell_i^j} \le 1$$ proposed in \cite{yao10}.
We can also generalize the entropy bound.

\begin{theorem}[Entropy Bound] \label{thm:entropy}
	Fix a positive real number $D$.
	Let $\mathcal{Q}$ be a uniquely decodable code for a source random variable $Z$ with probability $\{p_1, p_2, \ldots, p_m\}$ and $D$-ary entropy $H_D(Z)$.
	Then,
	\begin{equation} \label{eq:entropy}
		\sum_{j = 1}^m p_j \sum_{i = 1}^n \ell_i^j \log_D q_i \ge H_D(Z),
	\end{equation}
	where the equality holds if and only if $\sum_{i = 1}^n \ell_i^j \log_D q_i = -\log_D p_j$.
\end{theorem}

\begin{IEEEproof}
	See Appendix~\ref{sec:kraft_entropy}.
\end{IEEEproof}

The L.H.S. of \eqref{eq:entropy} looks a bit different from the traditional entropy bound or the multichannel version in \cite{yao10}.
Here we give an explanation on the meaning of this entropy bound.

We first consider the simple case of single channel source codes, \ie, $n = 1$.
Suppose we choose $D = q_1$, the left hand side of \eqref{eq:entropy} is the average codeword length.
let $s_j$ be the sum of the codeword lengths over all the channels of the $j$-th codeword, \ie, $s_j = \sum_{i = 1}^n \ell_i^j$.
The left hand side of \eqref{eq:entropy} is the average of $s_j$.
That is, we concatenate all the channels of a codeword together and treat it as a codeword in a single channel.

However, when we have heterogeneous alphabet sizes, the codeword length of each channel is in a different unit.
The length $(\ell_1^j, \ldots, \ell_n^j)$ means that the codeword in the $i$-th channel takes $\ell_i^j$ symbols from a $q_j$-ary alphabet.
To unify the measurements, we express $\ell_i^j$ symbols from a $q_i$-ary alphabet by using $\ell_i^j \log_D q_i$ $D$-ary symbols.
Then, the length of the $j$-th codeword after concatenating all the channels is $\sum_{i = 1}^n \ell_i^j \log_D q_i$ $D$-ary symbols. 
The average of this length among all the codewords is the L.H.S. of \eqref{eq:entropy}.

In practice, we want a codeword of a uniquely decodable code to be decoded without referring to the symbols of any future codewords.
A multichannel source code having this property is known as a \emph{strongly self-punctuating code} \cite{yao10}.

\begin{definition}[Strongly Self-Punctuating Codes]
	A source code $\mathcal{Q}$ is a \emph{strongly self-punctuating code} if, for any $n$ sequences $\mathcal{S}_i \subseteq \mathcal{Z}_i^\ast$, $i = 1, 2, \ldots, n$, there exists no more than one codeword in $\mathcal{Q}$ such that the $i$-th component of the codeword is a prefix of $\mathcal{S}_i$ for all $i = 1, 2, \ldots, n$.
\end{definition}

By convention, the multichannel prefix-free codes are also called the multichannel prefix codes.
By \cite[Thm.~1]{yao10},\footnote{Although \cite{yao10} only considered homogeneous alphabet size, the proof of Theorem~1 in \cite{yao10} is independent of alphabet sizes.
That is, the same proof is still valid for heterogeneous alphabet sizes.}
we know that a multichannel source code is a strongly self-punctuating code if and only if it is a prefix-free code.

\section{Rectangle Packing and Prefix-Free Codes}

A \emph{rectangle packing problem} is a decision procedure for the packability of a given set of two-dimensional parallel\footnote{%
	A set of rectangles are \emph{parallel} if all their edges are either parallel or orthogonal to each other.}
rectangular blocks with fixed orientations in a given enclosing two-dimensional container without any overlapping.

In this section, we show the relation between the existence of prefix codes with a given multiset of lengths and the existence of solutions of a special case of the rectangle packing problem.

Let $(\ell_1^j, \ldots, \ell_n^j)$ be the length of the $j$-th codeword in a codebook of size $m$.
Define $\ellmax_i := \max_{j = 1}^m \ell_i^j$ as the maximum length of all codewords in the $i$-th channel. 

\subsection{Single Channel Case}

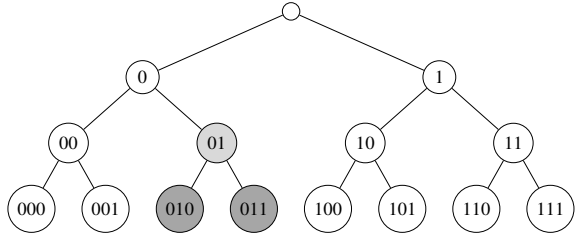
\begin{figure}
	\centering
	\begin{tikzpicture}[scale=.7,every tree node/.style={draw,circle},level distance=1.25cm,sibling distance=.5cm,edge from parent path={(\tikzparentnode)--(\tikzchildnode)}]
		\Tree [.{}
			[.0
				[.00
					[.000 ] [.001 ]
				]
				[.\node[fill=gray!30] {01};
					[.\node[fill=gray!70] {010}; ] [.\node[fill=gray!70] {011}; ]
				]
			]
			[.1
				[.10
					[.100 ] [.101 ]
				]
				[.11
					[.110 ] [.111 ]
				]
			]
		]
	\end{tikzpicture}
	\caption{An example of a binary prefix tree where the maximum codeword length is $3$.
	If $01$ is chosen as a codeword, then all its descendants, \ie, $010$ and $011$ in the figure, cannot be chosen as a codeword anymore.}
	\label{fig:prefix_tree}
\end{figure}

We first consider the single channel case, \ie, $n = 1$.
For brevity, we omit the subscripts denoting the $1$-st channel, which is the only channel we have in this subsection.
A prefix tree can be used to show the prefix relationship between the words in $\mathcal{Z}^\ast$.
Suppose the maximum codeword length $\ellmax$ is given, then the prefix tree is a complete $q$-ary tree of height $\ellmax$, \ie, the longest root-to-leaf path contains $\ellmax$ edges.
If a node in the tree is chosen as a codeword of a prefix code, then all its descendants cannot be chosen as a codeword anymore.
That is, a node is not prefix free to all its descendants.
Fig.~\ref{fig:prefix_tree} illustrates an example of a binary prefix tree.

As selecting a node makes all its descendants unavailable, it is convenient to view the selection as eliminating the selected node and all its descendants.
A word of length $\ell$ will eliminate $q^{\ellmax - \ell}$ consecutive leaves.
If we use a rectangle, or a \emph{block}, to represent the set of consecutive leaves to be eliminated after selecting a node, then the width of the block, \ie, the number of leaves being represented, must be a power of $q$.

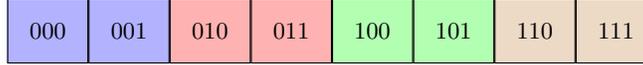
\begin{figure}
	\centering
	\small
	\begin{tikzpicture}
		\matrix (table) [matrix of math nodes,nodes={draw,minimum height=6ex,text width=6ex,align=center}] {
			000 & 001 & 010 & 011 & 100 & 101 & 110 & 111 \\
		};
		\begin{scope}[on background layer]
			\node[fill=red!30,inner xsep=0mm,inner ysep=0mm, fit=(table-1-3) (table-1-4) (table-1-4) (table-1-3)] {};
			\node[fill=blue!30,inner xsep=0mm,inner ysep=0mm, fit=(table-1-1) (table-1-2) (table-1-2) (table-1-1)] {};
			\node[fill=green!30,inner xsep=0mm,inner ysep=0mm, fit=(table-1-5) (table-1-6) (table-1-6) (table-1-5)] {};
			\node[fill=brown!30,inner xsep=0mm,inner ysep=0mm, fit=(table-1-7) (table-1-8) (table-1-8) (table-1-7)] {};
		\end{scope}
	\end{tikzpicture}
	\caption{The container containing all the leaf nodes of the prefix tree shown in Fig~\ref{fig:prefix_tree}.
	Each colored region corresponds to the leaf nodes of the subtree rooted at either $00, 01, 10$ or $11$.
	If a word of length $2$ is chosen as a codeword of a prefix-free code, then one of the colors will be selected and that colored region will be occupied.
	For example, if $01$ is chosen, then all the words with the prefix $01$ will be eliminated, \ie, red region is occupied.
	}
	\label{fig:1dtable}
\end{figure}

On the other hand, we can use an enclosing rectangle, or a \emph{container}, to represent the set of all leaves of the prefix tree.
As shown in Fig.~\ref{fig:prefix_tree}, each possibility of selecting a node corresponding to a word of length $\ell$ is equivalent to packing the corresponding block of width $q^{\ellmax-\ell}$ into the container.
Fig.~\ref{fig:1dtable} illustrates the container which corresponds to the prefix tree shown in Fig.~\ref{fig:prefix_tree}, where each colored region corresponds to the leaves of the subtree rooted at either $00, 01, 10$ or $11$.
If a word of length $2$ is chosen as a codeword of the prefix code, then a block of width $2^{3-2} = 2$ will be packed into one of the colored region.
For example, if $01$ is chosen, then all words with the prefix $01$ will be eliminated, \ie, the block is packed in the red region.
Observe that we cannot pack the block across two colors at the same time as their prefixes are different, \ie, we have an alignment constraint on the elimination.

The assignment of a codeword depends on the position the block is packed in the container.
If two blocks overlap with each other, then the two corresponding codewords are not prefix-free to each other.
So, there exists a single channel prefix code with the given codeword lengths if and only if the corresponding blocks can be packed into the container with no overlapping and satisfying the alignment constraint.

If the blocks can be packed into the container, then the sum of the widths of the blocks must be less than or equal to the width of the container, \ie,
\begin{equation} \label{eq:sum_kraft}
	\sum_{i = 1}^m q^{\ellmax - \ell^i} \le q^{\ellmax},
\end{equation}
which gives the single channel Kraft inequality $\sum_{i = 1}^m q^{-\ell^i} \le 1$.
On the other hand, suppose we know that the given codeword lengths satisfy the Kraft inequality, then the inequality \eqref{eq:sum_kraft} also holds.
Without loss of generality, let $\ell^1 \le \ell^2 \le \ldots \le \ell^m = \ellmax$.
Their corresponding blocks have width $q^{\ellmax - \ell^1} \ge q^{\ellmax - \ell^2} \ge \ldots \ge q^{\ellmax - \ell^m} = 1$.
First, we can put a block of width $q^{\ellmax - \ell^1}$ to the leftmost position of the container.
Then, we can put a block of width $q^{\ellmax - \ell^2}$ just on the right of the first.
We can repeat the procedure to pack in all the remaining blocks.
Note that this packing method satisfies the alignment constraint.
As there is no gap between the packed blocks, we can conclude that the blocks can be packed into the container.
This procedure also gives the codewords of the prefix code if it exists.
A similar way to relate a single channel prefix code to the Kraft inequality can be found in \cite{ye01}.

\subsection{Multichannel Case}

Using a similar model, an $n$-channel codeword can be considered as an $n$-dimensional block.
The container which contains all possible words is now $n$-dimensional, where the $i$-th dimension corresponds to the $i$-th channel.
To generalize the notion of width, a block or a container occupying $a_i$ units of space in the $i$-th dimension is said to have size $[a_1,\ldots,a_n]$.
We use parentheses and brackets to distinguish the length and size.
We sometimes call a block (resp. container) with size $[a_1,\ldots,a_n]$ an ``$a_1 \times a_2 \times \ldots \times a_n$ block (resp. container)''.
When $n = 2$, we call $a_1$ and $a_2$ the \emph{width} and \emph{height} respectively.

Given the lengths of the codewords, the sizes of the corresponding blocks and the container can be uniquely determined.
A codeword having length $(\ell_1^j, \ldots, \ell_n^j)$ corresponds to a block having size
	$$\left[q_1^{\ellmax_1 - \ell_1^j}, \ldots, q_n^{\ellmax_n - \ell_n^j}\right].$$
These blocks are to be packed into a container of size 
	$$\left[q_1^{\ellmax_1}, \ldots, q_n^{\ellmax_n}\right].$$
If the $i$-th channel is unused by all codewords, then we have $\ellmax_i = 0$, \ie, the $i$-th dimension of the container has length $1$.
It is equivalent to consider a lower dimensional container by removing the $i$-th channel.

Fig.~\ref{fig:2dtable} illustrates an example of a two-dimensional container.
If a projection is applied to the width (or the height) of the container in the figure, it becomes the container we have shown in Fig.~\ref{fig:1dtable} (or its transpose).
That is, the alignment constraint is satisfied dimension-wise.

\begin{figure}
	\centering
	\small
	\begin{tikzpicture}
		\matrix (table) [matrix of math nodes,nodes={draw,minimum height=6ex,text width=6ex,align=center}] {
			\substack{000\\000} & \substack{001\\000} & \substack{010\\000} & \substack{011\\000} & \substack{100\\000} & \substack{101\\000} & \substack{110\\000} & \substack{111\\000} \\
			\substack{000\\001} & \substack{001\\001} & \substack{010\\001} & \substack{011\\001} & \substack{100\\001} & \substack{101\\001} & \substack{110\\001} & \substack{111\\001} \\
			\substack{000\\010} & \substack{001\\010} & \substack{010\\010} & \substack{011\\010} & \substack{100\\010} & \substack{101\\010} & \substack{110\\010} & \substack{111\\010} \\
			\substack{000\\011} & \substack{001\\011} & \substack{010\\011} & \substack{011\\011} & \substack{100\\011} & \substack{101\\011} & \substack{110\\011} & \substack{111\\011} \\
			\substack{000\\100} & \substack{001\\100} & \substack{010\\100} & \substack{011\\100} & \substack{100\\100} & \substack{101\\100} & \substack{110\\100} & \substack{111\\100} \\
			\substack{000\\101} & \substack{001\\101} & \substack{010\\101} & \substack{011\\101} & \substack{100\\101} & \substack{101\\101} & \substack{110\\101} & \substack{111\\101} \\
			\substack{000\\110} & \substack{001\\110} & \substack{010\\110} & \substack{011\\110} & \substack{100\\110} & \substack{101\\110} & \substack{110\\110} & \substack{111\\110} \\
			\substack{000\\111} & \substack{001\\111} & \substack{010\\111} & \substack{011\\111} & \substack{100\\111} & \substack{101\\111} & \substack{110\\111} & \substack{111\\111} \\
		};
	\end{tikzpicture}
	\caption{
		An example of a two-dimensional container for a two-channel prefix code.
		The width and height correspond to the first and second channels respectively, where the maximum codeword lengths of both channels are $3$.
	}
	\label{fig:2dtable}
\end{figure}

The following theorem implies that the following are equivalent: 1) the existence of a multichannel prefix code with a given multiset of codeword lengths, and 2) the existence of a solution of a rectangle packing problem with dimension-wise alignment constraints with the corresponding multiset of block sizes.

\begin{theorem} \label{thm:transform}
	Two codewords are prefix free to each other if and only if their corresponding blocks do not overlap.
\end{theorem}

\begin{IEEEproof} 
	By Definition~\ref{def:prefix}, two codewords are prefix free to each others if there exists at least one channel in which the corresponding two components are prefix free to each others.
	Now, consider a projection of the container to one of the dimension.
	If the blocks overlap in that dimension, it means that they are not prefix free to each others in that corresponding channel.
	It is not possible for the blocks to overlap in all dimensions, or otherwise it contradicts that there exists at least one channel they are prefix free to each other.
	So, the projections of the two blocks do not overlap in at least one of the dimensions, which implies that the two blocks do not overlap in the container.

	Conversely, suppose the two blocks do not overlap in the container.
	Then, we have at least one dimension in which their projections do not overlap.
	That is, they are prefix free to each other in that corresponding channel.
\end{IEEEproof}

It is trivial to prove that the codeword lengths of any prefix code must satisfy Kraft inequality in the perspective of rectangle packing.
To see why, first note that by Theorem~\ref{thm:transform} it is possible to pack all the blocks of the corresponding sizes inside an appropriate container.
It then follows that the sum of volumes  of all blocks must be at most the total volume of the container, where the volume of a block (resp. container) is defined as the product of the components in the size of the block (resp. container).
That is,
\begin{equation*} 
	\sum_{j = 1}^m \prod_{i = 1}^n q_i^{\ellmax_i - \ell_i^j} \le \prod_{i = 1}^n q_i^{\ellmax_i},
\end{equation*}
or equivalently 
\begin{equation*} 
\sum_{j = 1}^m \prod_{i = 1}^n q_i^{- \ell_i^j} \le 1,
\end{equation*}
which is exactly the Kraft inequality \eqref{eq:kraft}.

The converse is unfortunately not true. That is, the Kraft inequality, which accounts only for volumes but not geometry, is not sufficient to show the existence of a prefix code. We show by giving a counterexample below.

Consider a two-channel binary code.
Suppose there are two codewords in the code and the codeword lengths are $(1, 0)$ and $(0, 1)$ respectively.
The size of the container is $2 \times 2$, and the corresponding block sizes are $1 \times 2$ and $2 \times 1$.
The area of the container and the sum of the areas of the blocks are both $4$, so the multichannel Kraft inequality is satisfied.
No matter which of the two possible choice to put a $1 \times 2$ block, what remains is a $1 \times 2$ region which is impossible to contain the remaining $2 \times 1$ block.
We thus conclude that prefix codes with the given codeword lengths do not exist.

\section{Efficient Decision Procedure for Two-Channel Prefix-Free Codes}

In this section, we present an efficient decision procedure for the existence of two-channel prefix codes via a constrained two-dimensional rectangle packing problem.

\subsection{Problem Formulation}

Our proposed algorithm in Section~\ref{sec:algo} will split the empty spaces of a container into multiple smaller containers, so we will formulate a model which consider more than one containers.
We first give some definitions which will be used in the remaining text.

\begin{definition}[{[Informal Overview]} Regions, Blocks, and Containers]
    A region $\region(x,y,w,h)$ is a rectangular subset of points on the Cartesian plane with width $w$ and height $h$, where the left and bottom borders are closed, and the right and top borders are open. The location $(x,y)$ of a region is defined as the coordinate of the lower-left corner of the region. The size of the region is denoted by $[w,h]$.
    
    We pay special attention to ``\emph{regular}'' and ``\emph{aligned}'' regions, which are concepts defined with respect to the arities $q_1$ and $q_2$. A region is \emph{regular} if $x$ is a power of $q_1$ and $y$ is a power of $q_2$. It is \emph{aligned} if $x$ is a multiple of $w$, and $y$ is a multiple of $h$.
    
    A containers $C$ is simply a region with a semantic meaning: It is meant to ``contain'' block(s).
    A block $B$ can be considered as a region with a variable location. We can ``pack'' a block by assigning it a location, denoted by $B(x,y)$.
\end{definition}

\begin{definition}[Regions]
    Fix integers $q_1, q_2 > 1$.
    For $x, y \in \ZZ$ and $w, h \in \NN$, the \emph{region} $R = \region(x,y,w,h)$ is defined as $\region(x,y,w,h) := [x, x+w) \times [y, y+h) \subseteq \RR^2$.
	The tuples $\location(R) := (x,y)$ and $\size(R) = [w,h]$ are called the \emph{location} and the \emph{size} of the region respectively.\footnote{We only need to consider nonnegative integers $x, y$ and positive integers $w, h$ in this paper.}
    Two regions $\region(x,y,w,h)$ and $\region(x',y',w',h')$ are said to \emph{overlap} with each other if $\region(x,y,w,h) \cap \region(x',y',w',h') \neq \emptyset$.
    The area of $R$ is defined as $\area(R) := wh$.
    The notation of size $\size(\cdot)$ is extended naturally to sets, \ie, if $S$ is a set of regions, then $\size(S) := \{\size(R): R \in S\}$.
\end{definition}
    
\begin{definition}[Regularity and Alignment]
    A size $[w,h]$ is said to be \emph{regular} (with respect to $q_1$ and $q_2$) if $w$ is a power of $q_1$ and $h$ is a power of $q_2$. A region is said to be regular if its size is regular.
    A region $\region(x,y,w,h)$ is said to be \emph{aligned} if $x$ is a multiple of $w$ and $y$ is a multiple of $h$.    
\end{definition}

To compare and hence sort two-dimensional sizes in the algorithm to be discussed in Section~\ref{sec:algo}, we define a partial ordering which compares both the width and the height equally, and and a total ordering which first compares the maximums of the width and the height, then the width, and finally the height. 
Formally, we give the following definitions.

\begin{definition}[Partial and Total Ordering of Sizes]
    \label{def:partial_order}
    \label{def:total_order}
    Let $s_1 = [w_1,h_1]$ and $s_2 = [w_2,h_2]$ be two distinct sizes. We define a partial ordering $\succ$ and a total ordering $>$ of sizes.
    We write $s_1 \succ s_2$ if $w_1 \geq w_2$ and $h_1 \geq h_2$.
    We write $s_1 > s_2$ if one of the following is satisfied:
    \begin{itemize}
        \item $\max\{w_1,h_1\} > \max\{w_2,h_2\}$
        \item $\max\{w_1,h_1\} = \max\{w_2,h_2\}$ and $w_1 > w_2$
        \item $\max\{w_1,h_1\} = \max\{w_2,h_2\}$ and $w_1 = w_2$ and $h_1 > h_2$
    \end{itemize}
    Note that if $s_1 \succ s_2$ then $s_1 > s_2$.
\end{definition}

\begin{example}
	The sizes $[8, 8]$, $[8,4]$, $[8,2]$, $[4, 8]$, $[2, 8]$, $[4, 2]$ are sorted in descending order.
\end{example}

In the following, we state a basic geometric fact about regular aligned regions.

\begin{lemma}
    \label{lemma:overlap}
    If $R_1$ and $R_2$ are two regular aligned regions with $R_2 \succeq R_1$, then either $R_2$ completely covers $R_1$, or they do not overlap. 
    More precisely, if $R_1 = \region(x_1,y_1,w_1,h_1)$ and $R_2 = \region(x_2,y_2,w_2,h_2)$ are regular and aligned, $w_2 \geq w_1$, and $h_2 \geq h_1$, then one of the following must be true:
    \begin{itemize}
        \item $R_1 \cap R_2 = R_1$
        \item $R_1 \cap R_2 = \emptyset$
    \end{itemize}
\end{lemma}

\begin{IEEEproof}
    The lemma is quite obvious. For the sake of completeness, we provide a proof for the case where the lower-left corner of $R_2$ overlaps with the upper-right corner of $R_1$.
    
    Since $R_1$ and $R_2$ are regular and aligned, 
    for $i \in \{1,2\}$, 
    let $w_i = q_1^{a_i}$ and $x_i = b_i w_i = b_i q_1^{a_i}$,
    for some non-negative integers $a_i$ and integers $b_i$.
    Since $w_2 \geq w_1$, we have $a_2 \geq a_1$.
    
    Suppose $R_1 \cap R_2 \neq \emptyset$. 
    Consider the case where the lower-left corner of $R_2$ overlaps with the upper-right corner of $R_1$. 
    In particular, we have:
    \begin{align*}
    x_1 \leq x_2 < x_1 + w_1 \leq x_2 + w_2 \\
    b_1 q_1^{a_1} \leq b_2 q_1^{a_2} < (b_1 + 1) q_1^{a_1} \\
    b_1 \leq b_2 q_1^{a_2 - a_1} < b_1 + 1 \\
    \end{align*}
    Since $b_2 q_1^{a_2 - a_1}$ is an integer, the third inequality forces $b_2 q_1^{a_2 - a_1} = b_1$, which implies that $x_1 = b_1 q_1^{a_1} = b_2 q_1^{a_2} = x_2$. Using a similar argument, we can show that $y_1 = y_2$.
    Thus $R_1 \cap R_2 = R_1$.
    
    The other three cases can be proven analogously.
\end{IEEEproof}

The following lemma is a single dimension version of Lemma~\ref{lemma:overlap}.

\begin{lemma}
    \label{lemma:overlap2}
	Let $X_1 = [x_1, x_1+w_1)$, $X_2 = [x_2, x_2+w_2)$, $Y_1 = [y_1, y_1+h_1)$ and $Y_2 = [y_2, y_2+h_2)$ be intervals where $w_1 \mid x_1$, $w_2 \mid x_2$, $h_1 \mid y_1$, $h_2 \mid y_2$, $w_1$ and $w_2$ are powers of $q_1$, $h_1$ and $h_2$ are powers of $q_2$, and $w_1 \le w_2$, $h_1 \le h_2$.
	Then, we have
	\begin{itemize}
		\item $X_1 \cap X_2 = X_1$ or $X_1 \cap X_2 = \emptyset$; and
		\item $Y_1 \cap Y_2 = Y_1$ or $Y_1 \cap Y_2 = \emptyset$.
	\end{itemize}
\end{lemma}

\begin{IEEEproof} 
	We can treat the intervals $X_1, X_2$ be a projection of regions.
	Let $R_1 = \region(x_1, 0, w_1, 1)$ and $R_2 = \region(x_2, 0, w_2, q_1)$.
	As we have $q_1 > 1$, thus $R_1 \preceq R_2$.
	Then, we can apply Lemma~\ref{lemma:overlap} to show that either $R_1 \cap R_2 = R_1$ or $R_1 \cap R_2 = \emptyset$, which corresponds to either $X_1 \cap X_2 = X_1$ or $X_1 \cap X_2 = \emptyset$.
	A similar argument can be used to prove the case for $Y_1$ and $Y_2$.
\end{IEEEproof}

\begin{definition}[Blocks and Containers]
	Let $[w,h]$ be a regular size.
	A \emph{block} $B_{w,h}$ is a function mapping locations to regions, \ie, $B_{w,h}: (u,v) \in \ZZ^2 \mapsto \region(u,v,w,h)$.
    The tuple $\size(B_{w,h}) := [w,h]$ is called the size of the block $B_{w,h}$.
    We omit the subscript in $B_{w,h}$ when the size is clear from the context.
    The value $\area(B) := wh$ is called the area of $B$.  

    A \emph{container} $C$ is simply a region $\region(x,y,w,h)$, for some $x,y \in \ZZ$ and $w,h \in \NN$, with semantic meaning. The definitions of the location, the size, and the area of a container is inherited from those of a region.
\end{definition}

Note that the sizes of the blocks transformed from the corresponding codeword lengths are always regular.

Consider a multiset\footnote{A multiset \cite{multiset} is a set with possibly repeated elements.} $\Bset = \{B_1,\ldots\}$ of regular blocks, and a set $\Cset = \{C_1, \ldots\}$ of non-overlapping containers.
Let $C_i = \region(x^i_C, y^i_C, w^i_C, h^i_C)$ be the $i$-th container, and $[w^i_B, h^i_B] = \size(B_i)$ be the size of the $i$-th block.

When we say some blocks or containers are sorted, it means that it is sorted by their sizes in descending order according to~\autoref{def:total_order}.
Without loss of generality, we assume $\Bset$ is sorted in \textbf{descending order}.

\begin{definition}[Constrained Rectangle Packing Problem]
    \label{def:problem}
    Let $\Bset = \{B_i\}_{i = 1}^{|\Bset|}$ be a multiset of blocks and $\Cset$ be a set of non-overlapping containers. The two-dimensional rectangle packing problem specified by $(\Bset,\Cset)$, is to find a simultaneous assignment $\{(x_i,y_i)\}_{i=1}^{|\Bset|}$ to all blocks in $\Bset$, such that the resulting regions $B(x_i,y_i)$ are aligned and non-overlapping, and each of them is contained in a container in $\Cset$, \ie, 
	\begin{equation*}
		\begin{cases}
        \forall B_i \in \Bset, B_i(x_i, y_i)~\text{is aligned}, \ie, w^i_B | x_i~\text{and}~h^i_B | y_i \\
		\forall B_i, B_j \in \Bset~\suchthat~i \neq j, B_i(x_i,y_i) \cap B_j(x_j,y_j) = \emptyset\\
		\forall B_i \in \Bset, \exists C \in \Cset~\suchthat~B_i(x_i,y_i) \subseteq C
		\end{cases}.
	\end{equation*}
    A simultaneous assignment $\{(x_i,y_i)\}_{i=1}^{|\Bset|}$ satisfying the above is called a solution. 
    If no such assignment exists, then we say that (the problem specified by) $(\Bset, \Cset)$ has no solution.  
    Note that if a solution exists, then each block $B_i$ must be contained in one and only one container, since the containers are non-overlapping.
\end{definition}

At the beginning when there is only one initial container, we have the problem $(\Bset,\Cset_0)$ where $\Cset_0 = \{C_0\}$, $C_0 = \region\left(0,0,q_1^{\ellmax_1},q_2^{\ell_2^\text{max}}\right)$.

\subsection{The Algorithm}\label{sec:algo}

Before stating our algorithm for the problem $(\Bset, \Cset)$, we introduce a sub-algorithm which aims to cut containers into smaller containers satisfying some constraints.
The cutting algorithm for the function defined as follows.

\begin{definition}[Container Cutting Function $\sigma$]
    \label{def:cutting_func}
    Let $C$ be a container and $s$ be a regular size.
    We model the cutting of containers by defining a function $\sigma(C, s)$ which maps the container $C$ and the size $s$ into the smallest set of non-overlapped regular aligned containers each with size $\preceq$ $s$.
    Also, the resulting containers must satisfy $\bigcup_{C' \in \sigma(C, s)} C' = C$.
    If no such set exists, then $\sigma(C, s) := \emptyset$, the empty set.
    The notation extends naturally to sets of non-overlapping containers.
    Let $\Cset$ be a set of containers. We define $\sigma(\Cset, s) := \bigcup_{C \in \Cset} \sigma(C, s)$.
\end{definition}

Note that the cutting function $\sigma$ might not be well defined since there may exist more than one smallest sets of such regions. To show that the cutting function is well defined (\cref{lem:cut_function}), we introduce the following notation.

\begin{definition}[Quotient Containers and Remainder Regions]
    Let $C = \region(x_C,y_C,w_C,h_C)$ be a container, and $[w,h]$ be a regular size. 
    The set of \emph{quotient containers} with respect to $[w,h]$ contained in $C$ is 
    the set of all regular aligned containers of size $[w,h]$ overlapping with $C$.
    Formally, it is defined as
    \[\Qset = \QuotientContainers(C,[w,h]) = \{\region(x,y,w,h) \subseteq C: w | x, h | y \}.\]

    In the typical case where $\Qset \neq \emptyset$, let $x_\Qset, y_\Qset, w_\Qset, h_\Qset$ be such that $\region(x_\Qset,y_\Qset,w_\Qset,h_\Qset) = \bigcup_{C' \in \Qset} C'$.
    Consider the space obtained by removing the region $\region(x_\Qset,y_\Qset,w_\Qset,h_\Qset)$ from $C$. 
    Such a space can be divided into a set of (at most) 8 \emph{remainder regions}, which is defined as
    \begin{align*}
    &\RemainderRegions(C,[w,h]) \\
    = 
    &\left \{ \begin{matrix}
    \region(x_\text{left},y_\text{high},w_\text{left},h_\text{high}), & \region(x_\text{mid},y_\text{high},w_\text{mid},h_\text{high}), & 
    \region(x_\text{right},y_\text{high},w_\text{right},h_\text{high}), \\
    \region(x_\text{left},y_\text{mid},w_\text{left},h_\text{mid}), & & \region(x_\text{right},y_\text{mid},w_\text{right},h_\text{mid}), \\
    \region(x_\text{left},y_\text{low},w_\text{left},h_\text{low}), & 
    \region(x_\text{mid},y_\text{low},w_\text{mid},h_\text{low}), &
    \region(x_\text{right},y_\text{low},w_\text{right},h_\text{low})
    \end{matrix}     \right \}
    \end{align*}
    where
    $x_\text{left} = x_C$, $x_\text{mid} = x_\Qset$, $x_\text{right} = x_\Qset + w_\Qset$, 
    $y_\text{low} = y_C$, $y_\text{mid} = y_\Qset$, $y_\text{high} = y_\Qset + h_\Qset$,
    $w_\text{left} = x_\Qset - x_C$, $w_\text{mid} = w_\Qset$, $w_\text{right} = (x_C + w_C) - (x_\Qset + w_\Qset)$, 
    $h_\text{low} = y_\Qset - y_C$, $h_\text{mid} = h_\Qset$, and $h_\text{high} = (y_C + h_C) - (y_\Qset + h_\Qset)$. 
    
    In the event that $\Qset = \emptyset$, then we define $\RemainderRegions(C,[w,h]) = C$.   Note that for any size $[w,h]$ it holds that $C = \QuotientContainers(C,[w,h]) \cup \RemainderRegions(C,[w,h])$.
\end{definition}

Note that there are other ways to divide the space obtained by removing the quotient containers from $C$ into regions. We define the remainder regions with the intuition that no regular aligned regions of size at most $[w,h]$ can be found across the boundaries of two remainder regions. Therefore it does not really matter how the remainder regions are defined. Formally, we prove the following lemma.

\begin{lemma}
    \label{lem:remainder}
    Let $C$ be a container, $[w,h]$ and $[w',h']$ be regular sizes with $[w,h] \succeq [w',h']$.
    Let $\Rset = \RemainderRegions(C, [w,h])$.
    Let $\Rset' \subseteq \Rset$ be any subset of remainder containers covering a rectangular space, \ie, one can define $x_R,y_R,w_R,h_R$ such that $R = \region(x_R,y_R,w_R,h_R) = \bigcup_{R' \in \Rset'} R'$.
    Then $$\QuotientContainers(R, [w',h']) = \bigcup_{R' \in \Rset'} \QuotientContainers(R', [w',h']).$$
\end{lemma}

\begin{IEEEproof} 
	Note that $|\Rset'| \le 3$ and the proof is trivial for $|\Rset'| < 2$.
	Also, it is obvious that $\QuotientContainers(R,[w',h'])$ and $\bigcup_{R' \in \Rset'} \QuotientContainers(R',[w',h'])$ are empty sets when $[w,h] = [w',h']$.
	If we have $|\Rset'| \ge 2$, then one of the corner in $\Rset$ must be in $\Rset'$.

	We first consider $|\Rset'| = 2$.
	Without loss of generality, let $\Rset'$ be the set containing the top-left and top-middle remainder regions, \ie,
	\begin{equation*}
		\Rset' = \{\region(x_\text{left},y_\text{high},w_\text{left},h_\text{high}), \region(x_\text{mid},y_\text{high},w_\text{mid},h_\text{high})\}.
	\end{equation*}
	The other configurations can be proved by similar arguments.

	It is clear that we have $\QuotientContainers(R, [w',h']) \supseteq \bigcup_{R' \in \Rset'} \QuotientContainers(R', [w',h'])$.

	Case I: $\QuotientContainers(R, [w',h']) = \emptyset$. The proof is done immediately.

	Case II: $\QuotientContainers(R, [w',h']) \neq \emptyset = \bigcup_{R' \in \Rset'} \QuotientContainers(R', [w',h'])$.
	Let $$Q = \bigcup_{Q' \in \QuotientContainers(R, [w',h'])} Q' = \region(x_Q, y_Q, w_Q, h_Q),$$ where $w' \mid x_Q$.
	We must have $w_Q < 2w'$, or otherwise $\bigcup_{R' \in \Rset'} \QuotientContainers(R', [w',h']) \neq \emptyset$.
	That is, we have $w_Q = w'$.
	Note that $Q \cap \region(x_\text{left},y_\text{high},w_\text{left},h_\text{high}) \neq \emptyset$ and $Q \cap \region(x_\text{mid},y_\text{high},w_\text{mid},h_\text{high}) \neq \emptyset$, or otherwise it leads to the same contradiction that $\bigcup_{R' \in \Rset'} \QuotientContainers(R', [w',h']) \neq \emptyset$.
	Let $Q \cap \region(x_\text{mid},y_\text{high},w_\text{mid},h_\text{high}) = \region(x_B, y_B, w_B, h_B)$, where $w \mid x_B > x_Q$.
	As $[w,h] \succeq [w',h']$ are both regular, we also have $w' \mid x_B$.
	On the other hand, we have $x_Q + w' > x_B > x_Q$ but $w' \mid x_Q, x_B$, which is a contradiction.
	That is, case II will not occur.

	Case III: $\QuotientContainers(R', [w',h']) \neq \emptyset$ for all $R' \in \Rset'$.
	Let the remainder regions of $\region(x_\text{left},y_\text{high},w_\text{left},h_\text{high})$ and $\region(x_\text{mid},y_\text{high},w_\text{mid},h_\text{high})$ be the sets
	\begin{equation*}
		\begin{Bmatrix}
			\region_{1A}, & \region_{2A}, & \region_{3A},\\
			\region_{4A}, & & \region_{6A},\\
			\region_{7A}, & \region_{8A}, & \region_{9A}
		\end{Bmatrix}
		\quad \text{ and } \quad
		\begin{Bmatrix}
			\region_{1B}, & \region_{2B}, & \region_{3B},\\
			\region_{4B}, & & \region_{6B},\\
			\region_{7B}, & \region_{8B}, & \region_{9B}
		\end{Bmatrix}
	\end{equation*}
	respectively in a geometric illustration.
	By the same means of illustration, we have
	\begin{equation*}
		\bigcup_{R' \in \Rset'} \RemainderRegions(R', [w',h']) = \begin{Bmatrix}
			\region_{1A}, & \region_{2A}, & \region_{3A}, & \region_{1B}, & \region_{2B}, & \region_{3B},\\
			\region_{4A}, & & \region_{6A}, & \region_{4B}, & & \region_{6B},\\
			\region_{7A}, & \region_{8A}, & \region_{9A}, & \region_{7B}, & \region_{8B}, & \region_{9B}
		\end{Bmatrix}
	\end{equation*}
	and
	\begin{equation*}
		\RemainderRegions(R, [w',h']) = \begin{Bmatrix}
			\region_{1A}, & \region_{2A} \cup \region_{3A} \cup \region_{1B} \cup \region_{2B}, & \region_{3B},\\
			\region_{4A}, & & \region_{6B},\\
			\region_{7A}, & \region_{8A} \cup \region_{9A} \cup \region_{7B} \cup \region_{8B}, & \region_{9B}
		\end{Bmatrix}.
	\end{equation*}
	Due to the fact that the height of $\region_i$ is less than $h'$ for $i = 2A, 3A, 1B, 2B, 8A, 9A, 7B, 8B$, we know that there is no subset of $\region_{2A} \cup \region_{3A} \cup \region_{1B} \cup \region_{2B}$ or $\region_{8A} \cup \region_{9A} \cup \region_{7B} \cup \region_{8B}$ is in $\QuotientContainers(R, [w',h'])$.
	Similarly, we do not need to consider $\region_i$ where $i = 1A, 4A, 7A, 3B, 6B, 9B$ as their width is less than $w'$.
	That is, we only need to show that $\region_{6A} \cup \region_{4B} = \emptyset$.

	For $i = 6A, 4B$, we write $\region_i = \region(x_i, y_i, w_i, h_i)$.
	Note that we have $w' \mid x_{6A}$, $w \mid x_{4B}$, $w' \mid x_{4B}+w_{4B}$, and $x_{6A}+w_{6A} = x_{4B}$, $0 \le w_{6A}, w_{4B} < w'$.
	So, we also have $0 \le w_{6A} + w_{4B} < 2w'$ and $w' \mid w_{6A} + w_{4B}$.
	This implies that $w_{6A} + w_{4B}$ is either $0$ or $w'$.
	Suppose $w_{6A} + w_{4B} = w'$, then we have $0 < w_{6A}, w_{4B} < w'$.
	As $[w,h] \succeq [w',h']$ are regular, we have $w' \mid x_{4B}$.
	That is, we have $x_{6A} + w' > x_{4B} > x_{6A}$ and $w' \mid x_{6A}, x_{4B}$, which is a contradiction.
	So, we must have $w_{6A} + w_{4B} = 0$.
	In other words, we have $\region_{6A} \cup \region_{4B} = \emptyset$.

	Case IV: there is one and only one $R' \in \Rset'$ such that $\QuotientContainers(R', [w',h']) = \emptyset$.
	Without loss of generality, let $\QuotientContainers(\region(x_\text{mid},y_\text{high},w_\text{mid},h_\text{high}), [w',h']) \neq \emptyset$.
	By using a similar argument in case III, we do not need to consider the upper, lower and the rightmost remainder regions.
	For the leftmost remainder region, first, we have $w \mid x_\text{mid}$.
	As $[w,h] \succeq [w',h']$ are regular, we also have $w' \mid x_\text{mid}$.
	So, we must have $w_\text{left} < w'$ or otherwise $\QuotientContainers(R', [w',h']) \neq \emptyset$ for both $R' \in \Rset'$.
	Then, we have $x_\text{left} = x_\text{mid} - w_\text{left}$ thus $w' \nmid x_\text{left}$, which means that $\QuotientContainers(R, [w',h']) = \bigcup_{R' \in \Rset'} \QuotientContainers(R', [w',h'])$.

	The above four cases finish the proof for $|\Rset'| = 2$.
	For $|\Rset'| = 3$, it can be done by simply applying twice the arguments for $|\Rset'| = 2$.
\end{IEEEproof}

With the above notions, we can now prove the uniqueness of the output of the cutting function.

\begin{lemma} \label{lem:cut_function}
    Let $\Cset$ be a set of non-overlapping containers and $[w,h]$ be a regular size.
    The set given by $\sigma(\Cset, [w, h])$ is unique, \ie, $\sigma$ is a function.
\end{lemma}

\begin{IEEEproof} 
    It suffices to show that $\sigma(C, [w, h])$ is unique for each $C \in \Cset$.
    Fix $C \in \Cset$ and let $C = \region(x_C,y_C,w_C,h_C)$.
    Let $\Sset$ be a smallest set of non-overlapping regular aligned containers each with width and height at most $w$ and $h$ respectively, and $C = \bigcup_{S \in \Sset} S$.
    We will show that such a set $\Sset$ is unique, and thus is the value of $\sigma(C, [w, h])$.
    Our argument is constructive.
    In particular, we will recursively determine disjoint subsets of $\Sset$, until the entire $\Sset$ is determined. This can be turned into an algorithm which computes $\sigma(C, [w, h])$.
    
    Let $[w',h']$ be a regular size with the largest width and height respectively such that $[w',h'] \preceq [w,h]$ and $\Qset = \QuotientContainers(C, [w',h']) \neq \emptyset$. 
	That is, $[w',h']$ is the largest size (in total order) of regular aligned regions we can cut from $\Cset$.
    We claim that $\Qset \subseteq \Sset$.
    Suppose not, then there exists a regular aligned container $C^* \in \Qset$ such that $C^* \notin \Sset$.
	By the definition of $[w',h']$, we know that every container in $\Sset$ has size $\preceq [w',h']$.
    By the definition of quotient containers, $C^* \subseteq C$.
    Since $C = \bigcup_{S \in \Sset} S$, we have $C^* \subseteq \bigcup_{S \in \Sset} S$. 
	So, there is a smallest subset $\Sset' \subseteq \Sset$ such that $C^* \subseteq \bigcup_{S \in \Sset'} S$.
	Note that $C^*$ and all $S \in \Sset'$ are regular aligned containers.
	For any $S \in \Sset'$, as we have $\size(S) \preceq [w',h'] = \size(C^*)$, we can apply \autoref{lemma:overlap} to show that either $S \cap C^* = S$ or $S \cap C^* = \emptyset$.
	The latter cannot be true as it contradicts the minimality of $\Sset'$.
	That is, we have $C^* = \bigcup_{S \in \Sset'} S \notin \Sset$, which contradicts the minimality of $\Sset$.
	So, we must have $C^* \in \Sset$.
    
    By the above, we have concluded that $\Qset \subseteq \Sset$. Now, consider the remainder regions $\Rset = \RemainderRegions(C, [w',h'])$. Note that by~\autoref{lem:remainder}, $\QuotientContainers$ acts locally on the remainder regions. We can thus apply the argument above recursively by substituting $C$ with $R$ for each $R \in \Rset$, and eventually uniquely determine $\Sset$.
\end{IEEEproof}

Building on the cutting algorithm, we propose a na\"ive greedy algorithm (Algorithm~\ref{alg:csp}) which solves $(\Bset, \Cset)$.
The algorithm cuts the containers at hand minimally into smaller regular aligned containers, such that one of the resulting containers is just enough to contain the largest unpacked block at hand. The algorithm naturally packs the largest block at hand to such a container, releases the unused space back to the set of containers at hand, and continues until all blocks are packed.

The na\"ive algorithm (Algorithm~\ref{alg:csp}) is very inefficient as it needs to compute at each step the container cutting function $\sigma$ entirely although only a small part of its output is used. Later, we will show an optimized algorithm which has input-output behavior identical to that of Algorithm~\ref{alg:csp} but is much more efficient by performing ``lazy'' computation of $\sigma$.

\begin{figure}
    \removelatexerror
    \begin{algorithm}[H]
        \caption{An algorithm solving $(\Bset, \Cset)$}
        \label{alg:csp}
        \KwData{Integers $q_1, q_2 > 1$ with respect to which regularity and alignment are defined; A multiset $\Bset = \{B_i\}_{i=1}^{|\Bset|}$ of regular blocks with sizes sorted in descending order; A set $\Cset$ of non-overlapping containers.}
        \KwResult{A solution to $(\Bset, \Cset)$ if one exists, or $\bot$ otherwise.}
        \For{$i \in [|\Bset|]$}{
            $s^* := \min\{s: \forall B \in \Bset,~s \succeq \size(B) \}$  \;
            \tcc{Let $s^*$ be the smallest size which is sufficient to cover each block in $\Bset$.}
            $\Cset \gets \sigma(\Cset, s^*)$ \;
            \tcc{Cut the containers in $\Cset$ into the smallest set of regular aligned containers each with size at most $s^*$.}
            $\Sset := \{C \in \Cset: \size(C) \succeq \size(B_i)\}$ \;
            \tcc{Consider the subset of containers which are large enough to contain $B_i$ which is the largest block at hand.} 
            \If{$\Sset = \emptyset$}{\Return $\bot$ \;}
            \tcc{If no such subset exists, the algorithm reports a failure.}
            $\bar \Sset := \argmin_{C \in \Sset} \{\size(C)\}$ \;
            \tcc{Among those containers, consider a subset which have the smallest size.}
            $C^* \gets_{\$} \bar \Sset$ \;
            \tcc{Choose an arbitrary container $C^*$ from this subset.}
            $\location_i := \location(C^*)$ \;
            $\Bset := \Bset \setminus \{B_i\}$ \;
            $\Cset := \Cset \setminus \{C^*\} \cup \{C^* \setminus B_i(\location_i)\}$ \;
            \tcc{Assign $B_i$ to the location of $C^*$ so that their lower-left corners overlap. Remove $B_i$ from the set of blocks at hand. Remove the space occupied by $B_i$ from the container $C^*$.}   
        }
    \Return $\{\location_i\}_{i=1}^{|\Bset|}$
    \end{algorithm}
    \vskip -.5cm
\end{figure}

\begin{theorem}\label{thm:main}
    Algorithm~\ref{alg:csp} outputs a solution to $(\Bset, \Cset)$ if and only if $(\Bset, \Cset)$ has a solution.
    Furthermore, Algorithm~\ref{alg:csp} is a decision procedure for $(\Bset, \Cset)$ if we modify it a little bit that
\begin{enumerate}[i)]
	\item output $0$ if Algorithm~\ref{alg:csp} outputs $\bot$; and
    \item output $1$ if Algorithm~\ref{alg:csp} does not output $\bot$.
\end{enumerate}
\end{theorem}

The problem of determining the existence of a solution to $(\Bset, \Cset)$ is a yes-no question.
It is obvious that the modifications for the decision procedure will not affect the correctness of the algorithm.

We devote the next two subsections to prove the above theorem.

\subsection{Properties of the Solutions}

In order to prove Theorem~\ref{thm:main}, we need to establish some properties regarding the solutions of $(\Bset, \Cset)$.

\begin{lemma} \label{lem:cut}
    Let $\Bset = \{B_1,\ldots,B_m\}$ be a multiset of blocks with sizes sorted in descending order, and $\Cset$ be a set of non-overlapping containers.
	Let $w_{max} = \max_{i \in [m]} w^i_B$ and $h_{max} = \max_{i \in [m]} h^i_B$ be the width and height of the widest and tallest blocks respectively.
	It holds that $(\Bset, \Cset)$ has a solution if and only if $(\Bset, \sigma(\Cset, [w_{max}, h_{max}]))$ has a solution.
    Furthermore, any solution to $(\Bset, \Cset)$ is also a solution to $(\Bset, \sigma(\Cset, s^*))$, vice versa.
\end{lemma}

\begin{IEEEproof}
    We first show that $(\Bset, \Cset)$ has a solution if $(\Bset, \sigma(\Cset, [w_{max}, h_{max}]))$ has a solution.
	Suppose the latter holds.
    By~\cref{def:problem}, there exists a simultaneous assignment $\{(x_i,y_i)\}_{i=1}^{|\Bset|}$ which satisfies the following
    \begin{equation*}
    \begin{cases}
    \forall B_i, B_j \in \Bset~\suchthat~i \neq j, B_i(x_i,y_i) \cap B_j(x_j,y_j) = \emptyset,\\
    \forall B_i \in \Bset, \exists C_i \in \sigma(\Cset, [w_{max}, h_{max}])~\suchthat~B_i(x_i,y_i) \subseteq C_i, \\
    \forall B_i \in \Bset, B_i(x_i, y_i)~\text{is aligned}, \ie, w^i_B | x_i~\text{and}~h^i_B | y_i.
    \end{cases}
    \end{equation*}
    Fix $i \in [|\Bset|]$.
    By the definition of $\sigma$ (\cref{def:cutting_func}), $\sigma(\Cset, [w_{max}, h_{max}]) = \bigcup_{C \in \Cset} \sigma(C, [w_{max}, h_{max}])$.
    Therefore since $C_i \in \sigma(\Cset, [w_{max}, h_{max}])$, there must exist $C^*_i \in \Cset$ such that $C_i \in \sigma(C^*_i, [w_{max}, h_{max}])$.
    Using~\cref{def:cutting_func} again, $C^*_i = \bigcup_{C' \in \sigma(C^*_i, [w_{max}, h_{max}])} C'$. Therefore $C_i \subseteq C^*_i$.
    From these observations, the second constraint of a solution implies that 
    \[
        \forall B_i \in \Bset, \exists C^*_i \in \Cset~\suchthat~B_i(x_i,y_i) \subseteq C^*_i
    \]
    which together with the first and the third constraints implies that $\{(x_i,y_i)\}_{i=1}^{|\Bset|}$ is a solution to $(\Bset, \Cset)$.

    Next, we show that $(\Bset, \Cset)$ has a solution only if $(\Bset, \sigma(\Cset, [w_{max}, h_{max}]))$ has a solution. 
    By~\cref{def:problem}, since $(\Bset, \Cset)$ has a solution, there exists a simultaneous assignment $\{(x_i,y_i)\}_{i=1}^{|\Bset|}$ such that which satisfies the following
    \begin{equation*}
        \begin{cases}
            \forall B_i, B_j \in \Bset~\suchthat~i \neq j, B_i(x_i,y_i) \cap B_j(x_j,y_j) = \emptyset,\\
            \forall B_i \in \Bset, \exists C_i \in \Cset~\suchthat~B_i(x_i,y_i) \subseteq C_i, \\
            \forall B_i \in \Bset, B_i(x_i, y_i)~\text{is aligned}, \ie, w^i_B | x_i~\text{and}~h^i_B | y_i.
        \end{cases}
    \end{equation*}

	We have to show that for any $B_i(x_i,y_i)$, it must be contained in one of the containers in $\sigma(\Cset,[w_{max}, h_{max}])$.
	First, we know that $B_i(x_i,y_i) \subseteq C_i = \bigcup_{C \in \sigma(C_i, [w_{max}, h_{max}])} C$.
	So, there exists at least one container $\bar{C} \in \sigma(C_i, [w_{max}, h_{max}])$ such that $\bar{C} \cap B_i(x_i,y_i) \neq \emptyset$.
	Let $\region(x_C, y_C, w_C, h_C) := \bar{C}$ and $\region(x_i, y_i, w_i, h_i) := B_i(x_i, y_i)$.

	Case I: $\size(\bar{C}) \succeq \size(B_i)$.
	By Lemma~\ref{lemma:overlap}, we have $B_i(x_i,y_i) \subseteq \bar{C}$.

	Case II: $w_C \le w_B$ and $h_C > h_B$.
	Without loss of generality, let $\bar{C}$ be the tallest container in $\sigma(C_i, [w_{max}, h_{max}])$ which overlaps with $B_i(x_i,y_i)$.
	By Lemma~\ref{lemma:overlap2}, we have $x_i \le x_C < x_C + w_C \le x_i+w_i$, and $y_C \le y_i < y_i + h_i \le y_C + h_C$.
	Suppose there is a container $R = \region(x, y, w, h)$ overlaps the region $\region(x_i, y_C, x_C-x_i, h_C)$.
	By Lemma~\ref{lemma:overlap2}, we have $x \le x_i < x_i + w_i \le x+w$ if $w \ge w_i$, or $x_i \le x < x+w \le x_i+w_i$ if $w \le w_i$.
	The former case $R$ overlaps with $\bar{C}$, thus we must have the latter case with a refinement that $x_i \le x < x+w \le x_C$.
	It is similar for the a container which overlaps the region $\region(x_C+w_C, y_C, (x_i+w_i)-(x_C+w_C), h_C)$.
	That is, any container convering the region $\region(x_i, y_C, w_i, h_C)$ must stay in the interval $[x_i,x_i+w_i)$ for the width dimension.

	Now, we want to find a smallest set of containers in $\sigma(C_i, [w_{max}, h_{max}])$ covering the region $\region(x_i, y_C, x_C-x_i, h_C)$.
	As shown above, those containers in the set must stay in the interval $[x_i,x_C)$ for the width dimension.
	We cannot have a container in the set taller than $\bar{C}$ as this container must overlap with $B_i(x_i,y_i)$, which contradicts that $\bar{C}$ is the tallest.
	That is, every container in the set has size $\preceq [x_C-x_i, h_C]$.
	By Lemma~\ref{lemma:overlap}, every container in the set is a contained by $\region(x_i, y_C, x_C-x_i, h_C)$.
	So, we have a set of containers in $\sigma(C_i, [w_{max}, h_{max}])$ where the union of all containers in it is the regular aligned region $\region(x_i, y_C, x_C-x_i, h_C)$, which contradicts the minimality of $\sigma$.

	Case III: $w_C > w_B$ and $h_C \ge h_B$.
	A similar argument to the one for case II can be applied to show that this case cannot happen.

	Case IV: $\size(\bar{C}) \preceq \size(B_i)$.
	If there exists another container in $\sigma(C_i, [w_{max}, h_{max}])$ which overlaps with $B_i(x_i,y_i)$ such that it is wider or taller than $B_i(x_i, y_i)$, we would choose that container as $\bar{C}$ and go to case I, II or III.
	Here, we consider every container in $\sigma(C_i, [w_{max}, h_{max}])$ which overlaps with $B_i(x_i,y_i)$ has size $\preceq \size(B_i)$, and by Lemma~\ref{lemma:overlap}, $B_i(x_i,y_i)$ contains all these containers.
	That is, their union is exactly the regular aligned region $B_i(x_i,y_i)$, which contradicts the minimality of $\sigma$.

	The above four cases conclude that after the cut by $\sigma$, any block $B_i(x_i,y_i)$ must be contained in one of the containers in $\sigma(C_i, [w_{max}, h_{max}])$, which implies that $(\Bset, \sigma(\Cset, [w_{max}, h_{max}])$ has a solution.

	Note that we did not change the assignments of blocks during the whole proof, which means that any solution to $(\Bset, \Cset)$ is also a solution to $(\Bset, \sigma(\Cset, s^*))$, vice versa.
\end{IEEEproof}

In Lemma~\ref{lem:cut}, $B_1$ is the largest unpacked block at hand, so the width or the height of $\size(B_1)$ is the longest among the unpacked blocks.
This means that the longer side of $\size(B_1)$ equals to the corresponding side in $s^\ast$, \ie, the longer side of $\size(B_1)$ must fit the corresponding side of every container in $S$.
Then, the problem becomes a one-dimensional packing problem for the block.

\begin{lemma} \label{lem:swap}
    Let $\Bset$ be a multiset of regular blocks.
    Let $\Cset$ be a set of non-overlapping regular aligned containers such that there exist $C, C' \in \Cset$ with $\size(C) = \size(C')$.
    Suppose $(\Bset, \Cset)$ has a solution $\{(x_i,y_i)\}_{i = 1}^{|\Bset|}$. Then $\{(x'_i,y'_i)\}_{i = 1}^{|\Bset|}$ defined below is also a solution to $(\Bset, \Cset)$.
    
    \begin{align*}
        (x'_i, y'_i) = 
        \begin{cases}
        (x_i, y_i) - \location(C) + \location(C') & B_i(x_i, y_i) \subseteq C \\
        (x_i, y_i) - \location(C') + \location(C) & B_i(x_i, y_i) \subseteq C' \\
        (x_i, y_i) & \text{otherwise}
        \end{cases}
    \end{align*}
    where operations on tuples are performed coordinate-wise.
\end{lemma}

\begin{IEEEproof}
	For every block $B_i$ where $[w_i, h_i] := \size(B_i)$ and $B_i(x_i,y_i) \subseteq C$, we have $w_i \mid x_i$ and $h_i \mid y_i$ as $\{(x_i,y_i)\}_{i = 1}^{|\Bset|}$ is a solution to $(\Bset, \Cset)$.
	We must have $\size(B_i) \preceq \size(C)$ or otherwise $C$ is not largr enough to contain $B_i$.
	Then, we have $w_i \mid x_C$ and $h_i \mid y_C$ where $(x_C, y_C) := \location(C)$.
	As $\size(C) = \size(C')$, we also have $w_i \mid x_{C'}$ and $h_i \mid y_{C'}$ where $(x_C, y_C) = \location(C')$.
	So, we have $w_i \mid (x_i - x_C + x_{C'})$ and $h_i \mid (y_i - y_C + y_{C'})$.
	On the other hand, we have $x_C \le x_i+w_i \le x_C+w_C$ and $y_C \le y_i+h_i \le y_C+h_C$ where $[w_C, h_C] := \size(C) = \size(C')$.
	So, we have $x_{C'} \le x'_i + w_i \le x_{C'} + w_C$ and $y_{C'} \le y'_i + h_i \le y_{C'} + h_C$.
	That is, $B_i(x'_i, y'_i) \subseteq C'$ is a regular aligned region.
	
	All the blocks packed in a container is non-overlapping, thus after moving all the blocks in $C$ to $C'$, they are still non-overlapping.
	This means that we can swap the blocks contained in $C$ and $C'$ and obtain another solution.
\end{IEEEproof}

\begin{figure}
	\centering
	\begin{tikzpicture}[scale=.9]
		\fill[red!40!white] (0,3) rectangle (2,4);
		\fill[blue!40!white] (0,2) rectangle (2,3);
		\fill[green!40!white] (0,0) rectangle (2,2);

		\draw[gray!40!white] (0,0) rectangle (1,1);
		\draw[gray!40!white] (1,0) rectangle (1.5,1);
		\draw[gray!40!white] (1.5,0) rectangle (2,.5);
		\draw[gray!40!white] (1.5,.5) rectangle (2,1);
		\draw[gray!40!white] (0,1) rectangle (1,1.5);
		\draw[gray!40!white] (0,1.5) rectangle (.5,2);
		\draw[gray!40!white] (.5,1.5) rectangle (1,2);
		\draw[gray!40!white] (1,1) rectangle (1.5,1.5);
		\draw[gray!40!white] (1,1.5) rectangle (1.5,2);
		\draw[gray!40!white] (1.5,1) rectangle (2,2);

		\draw[gray!40!white] (0,2) rectangle (2,2.5);
		\draw[gray!40!white] (0,2.5) rectangle (1,3);
		\draw[gray!40!white] (1,2.5) rectangle (1.5,3);
		\draw[gray!40!white] (1.5,2.5) rectangle (2,3);

		\draw[gray!40!white] (0,3) rectangle (2,4);

		\draw[dashed] (0,2) -- (2,2);
		\draw (0,0) rectangle (2,4);

		\draw[thick,->] (2.1,2) -- (2.9,2);

		\fill[red!40!white] (3,1) rectangle (5,2);
		\fill[blue!40!white] (3,0) rectangle (5,1);
		\fill[green!40!white] (3,2) rectangle (5,4);

		\draw[gray!40!white] (3,2) rectangle (4,3);
		\draw[gray!40!white] (4,2) rectangle (4.5,3);
		\draw[gray!40!white] (4.5,2) rectangle (5,2.5);
		\draw[gray!40!white] (4.5,2.5) rectangle (5,3);
		\draw[gray!40!white] (3,3) rectangle (4,3.5);
		\draw[gray!40!white] (3,3.5) rectangle (3.5,4);
		\draw[gray!40!white] (3.5,3.5) rectangle (4,4);
		\draw[gray!40!white] (4,3) rectangle (4.5,3.5);
		\draw[gray!40!white] (4,3.5) rectangle (4.5,4);
		\draw[gray!40!white] (4.5,3) rectangle (5,4);

		\draw[gray!40!white] (3,0) rectangle (5,.5);
		\draw[gray!40!white] (3,.5) rectangle (4,1);
		\draw[gray!40!white] (4,.5) rectangle (4.5,1);
		\draw[gray!40!white] (4.5,.5) rectangle (5,1);

		\draw[gray!40!white] (3,1) rectangle (5,2);

		\draw (3,2) -- (5,2);
		\draw[dashed] (3,1) -- (5,1);
		\draw (3,0) rectangle (5,4);

		\draw[thick,->] (5.1,2) -- (5.9,2);

		\fill[red!40!white] (6,0) rectangle (8,1);
		\fill[blue!40!white] (6,1) rectangle (8,2);
		\fill[green!40!white] (6,2) rectangle (8,4);

		\draw[gray!40!white] (6,2) rectangle (7,3);
		\draw[gray!40!white] (7,2) rectangle (7.5,3);
		\draw[gray!40!white] (7.5,2) rectangle (8,2.5);
		\draw[gray!40!white] (7.5,2.5) rectangle (8,3);
		\draw[gray!40!white] (6,3) rectangle (7,3.5);
		\draw[gray!40!white] (6,3.5) rectangle (6.5,4);
		\draw[gray!40!white] (6.5,3.5) rectangle (7,4);
		\draw[gray!40!white] (7,3) rectangle (7.5,3.5);
		\draw[gray!40!white] (7,3.5) rectangle (7.5,4);
		\draw[gray!40!white] (7.5,3) rectangle (8,4);

		\draw[gray!40!white] (6,1) rectangle (8,1.5);
		\draw[gray!40!white] (6,1.5) rectangle (7,2);
		\draw[gray!40!white] (7,1.5) rectangle (7.5,2);
		\draw[gray!40!white] (7.5,1.5) rectangle (8,2);

		\draw[gray!40!white] (6,0) rectangle (8,1);

		\draw (6,2) -- (8,2);
		\draw (6,1) -- (8,1);
		\draw (6,0) rectangle (8,4);

	\end{tikzpicture}
	\caption{
		An example on the reason why we can put the largest block at the bottom left corner.
		By Lemma~\ref{lemma:overlap}, no block in the solution can cross the boundary shown by the dashed line.
		So, we can swap the upper half (red and blue) and the lower half (green) and obtain another solution.
		A similar argument can be used repeatedly until the red block is put at the bottom.
	}
	\label{fig:lowest}
\end{figure}
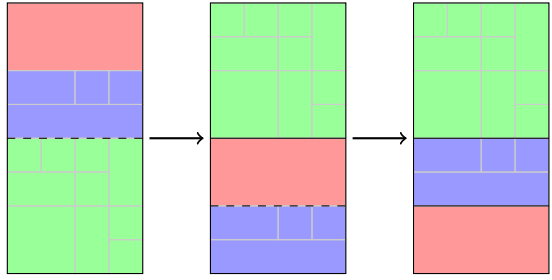

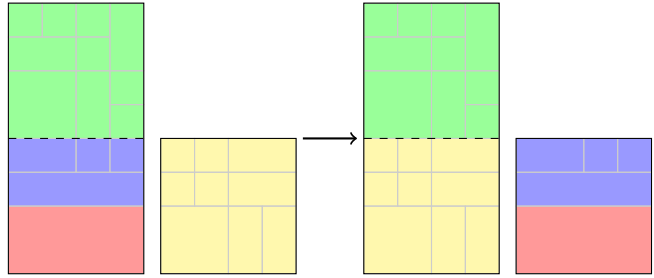
\begin{figure}
	\centering
	\begin{tikzpicture}[scale=.9]
		\fill[red!40!white] (0,0) rectangle (2,1);
		\fill[blue!40!white] (0,1) rectangle (2,2);
		\fill[green!40!white] (0,2) rectangle (2,4);

		\draw[gray!40!white] (0,2) rectangle (1,3);
		\draw[gray!40!white] (1,2) rectangle (1.5,3);
		\draw[gray!40!white] (1.5,2) rectangle (2,2.5);
		\draw[gray!40!white] (1.5,2.5) rectangle (2,3);
		\draw[gray!40!white] (0,3) rectangle (1,3.5);
		\draw[gray!40!white] (0,3.5) rectangle (0.5,4);
		\draw[gray!40!white] (0.5,3.5) rectangle (1,4);
		\draw[gray!40!white] (1,3) rectangle (1.5,3.5);
		\draw[gray!40!white] (1,3.5) rectangle (1.5,4);
		\draw[gray!40!white] (1.5,3) rectangle (2,4);

		\draw[gray!40!white] (0,1) rectangle (2,1.5);
		\draw[gray!40!white] (0,1.5) rectangle (1,2);
		\draw[gray!40!white] (1,1.5) rectangle (1.5,2);
		\draw[gray!40!white] (1.5,1.5) rectangle (2,2);

		\draw[gray!40!white] (0,0) rectangle (2,1);

		\draw[dashed] (0,2) -- (2,2);
		\draw (0,0) rectangle (2,4);

		\fill[yellow!40!white] (2.25,0) rectangle (4.25,2);

		\draw[gray!40!white] (2.25,0) rectangle (3.25,1);
		\draw[gray!40!white] (3.25,0) rectangle (3.75,1);
		\draw[gray!40!white] (3.75,0) rectangle (4.25,1);
		\draw[gray!40!white] (2.25,1) rectangle (2.75,1.5);
		\draw[gray!40!white] (2.75,1) rectangle (3.25,1.5);
		\draw[gray!40!white] (2.25,1.5) rectangle (2.75,2);
		\draw[gray!40!white] (2.75,1.5) rectangle (3.25,2);
		\draw[gray!40!white] (3.25,1) rectangle (4.25,1.5);
		\draw[gray!40!white] (3.25,1.5) rectangle (4.25,2);

		\draw (2.25,0) rectangle (4.25,2);

		\draw[thick,->] (4.35,2) -- (5.15,2);

		\fill[yellow!40!white] (5.25,0) rectangle (7.25,2);
		\fill[green!40!white] (5.25,2) rectangle (7.25,4);

		\draw[gray!40!white] (5.25,2) rectangle (6.25,3);
		\draw[gray!40!white] (6.25,2) rectangle (6.75,3);
		\draw[gray!40!white] (6.75,2) rectangle (7.25,2.5);
		\draw[gray!40!white] (6.75,2.5) rectangle (7.25,3);
		\draw[gray!40!white] (5.25,3) rectangle (6.25,3.5);
		\draw[gray!40!white] (5.25,3.5) rectangle (5.75,4);
		\draw[gray!40!white] (5.75,3.5) rectangle (6.25,4);
		\draw[gray!40!white] (6.25,3) rectangle (6.75,3.5);
		\draw[gray!40!white] (6.25,3.5) rectangle (6.75,4);
		\draw[gray!40!white] (6.75,3) rectangle (7.25,4);

		\draw[gray!40!white] (5.25,0) rectangle (6.25,1);
		\draw[gray!40!white] (6.25,0) rectangle (6.75,1);
		\draw[gray!40!white] (6.75,0) rectangle (7.25,1);
		\draw[gray!40!white] (5.25,1) rectangle (5.75,1.5);
		\draw[gray!40!white] (5.75,1) rectangle (6.25,1.5);
		\draw[gray!40!white] (5.25,1.5) rectangle (5.75,2);
		\draw[gray!40!white] (5.75,1.5) rectangle (6.25,2);
		\draw[gray!40!white] (6.25,1) rectangle (7.25,1.5);
		\draw[gray!40!white] (6.25,1.5) rectangle (7.25,2);

		\draw[dashed] (5.25,2) -- (7.25,2);
		\draw (5.25,0) rectangle (7.25,4);

		\fill[red!40!white] (7.5,0) rectangle (9.5,1);
		\fill[blue!40!white] (7.5,1) rectangle (9.5,2);

		\draw[gray!40!white] (7.5,1) rectangle (9.5,1.5);
		\draw[gray!40!white] (7.5,1.5) rectangle (8.5,2);
		\draw[gray!40!white] (8.5,1.5) rectangle (9,2);
		\draw[gray!40!white] (9,1.5) rectangle (9.5,2);

		\draw[gray!40!white] (7.5,0) rectangle (9.5,1);

		\draw (7.5,0) rectangle (9.5,2);

	\end{tikzpicture}
	\caption{
		An example on the reason why can select a smallest container in $\Sset$, which is the yellow container in the figure.
		Suppose there is a solution that the red block is put in a larger container.
		We can first swap the red block to the bottom as shown in Fig.~\ref{fig:lowest}.
		The size of the region below the dashed line is the same as the size of the yellow container.
		By Lemma~\ref{lemma:overlap}, no block in the solution can cross the boundary shown by the dashed line.
		So, we can simply swap all the blocks below the dashed line with those in the yellow container.
	}
	\label{fig:smallest}
\end{figure}

\begin{lemma} \label{lem:core}
    Let $\Bset = \{B_i\}_{i=1}^{|\Bset|}$ be a multiset of regular blocks with sizes sorted in descending order, and $\Cset$ be a set of non-overlapping containers.
    Suppose $(\Bset, \Cset)$ has a solution.
    Let $\Bset'$ and $\Cset'$ be the value of $\Bset$ and $\Cset$ after the first $(i=1)$ iteration. More precisely, let
    \begin{align*}
    \hat s := &\min\{s: \forall B \in \Bset,~s\succeq \size(B)\} \\
    \Sset := &\{C \in \sigma(\Cset, \hat s): \size(C) \succeq \size(B_1)\} \\
    C^* \in &\argmin_{C \in \Sset} \{\size(C)\} \\
    \Bset' := &\Bset \setminus \{B_1\} \\
    \Cset' := &\Cset \setminus \{C^*\} \cup \{C^* \setminus B_1(\location(C^*))\}
    \end{align*}
    Then $(\Bset',\Cset')$ also has a solution.
\end{lemma}

\begin{IEEEproof}
    By~\cref{lem:cut}, $(\Bset, \Cset)$ has a solution if and only if $(\Bset, \sigma(\Cset, \hat s))$ has a solution. We can thus assume that $(\Bset, \sigma(\Cset, \hat s))$ has a solution.
    We first argue that there must exist a solution $\{(x^*_i,y^*_i)\}_{i=1}^{|\Bset|}$ to $(\Bset, \sigma(\Cset, \hat s))$ with $B_1(x^*_1,y^*_1) \subseteq C^*$ (\cref{claim:claim1}).
    
    With the above claim, it remains to prove that it is fine to pack $B_1$ at the lower-left corner of $C^*$. Formally, we argue that if there exists a solution $\{(x^*_i,y^*_i)\}_{i=1}^{|\Bset|}$ to $(\Bset, \sigma(\Cset,\hat s))$ with $B_1(x^*_1,y^*_1) \subseteq C^*$, then there also exists a solution $\{(x^\dagger_i,y^\dagger_i)\}_{i=1}^{|\Bset|}$ to $(\Bset, \sigma(\Cset,\hat s))$ with $(x^\dagger_1, y^\dagger_1) = \location(C^*)$ (\cref{claim:claim2}). Once this is shown, the truthfulness of the lemma can then be verified by inspection.
\end{IEEEproof}
    
    \begin{claim}\label{claim:claim1}
        If $(\Bset, \sigma(\Cset, \hat s))$ has a solution, then there exists a solution $\{(x^*_i,y^*_i)\}_{i=1}^{|\Bset|}$ to $(\Bset, \sigma(\Cset, \hat s))$ with $B_1(x^*_1,y^*_1) \subseteq C^*$.
    \end{claim}
    
    \begin{IEEEproof}
    Let $\{(x_i,y_i)\}_{i=1}^{|\Bset|}$ be a solution to $(\Bset, \sigma(\Cset, \hat s))$. There exists $\bar C \in \Sset \subseteq \sigma(\Cset, \hat s)$ such that $B_1(x_1,y_1) \subseteq \bar C$. 
    Suppose $\bar C \in \argmin_{C \in \Sset} \{\size(C)\}$, then since $\size(\bar C) = \size(C^*)$, we can use~\cref{lem:swap} and construct a solution $\{(x^*_i,y^*_i)\}_{i=1}^{|\Bset|}$ with $B_1(x^*_1,y^*_1) \subseteq C^*$.
    
    Suppose otherwise $\bar C \in \Sset \setminus \argmin_{C \in \Sset} \{\size(C)\}$.
    Parse $\hat s$ as $\hat s = [\hat w,\hat h]$. 
    Since $B_1$ is the largest block in $\Bset$, we have $w_1 = \hat w$ or $h_1 = \hat h$. 
    We consider the case $w_1 = \hat w$. Similar arguments can be applied to the case $h_1 = \hat h$. Let $\bar s = [\bar w, \bar h] = \size(\bar C)$, and $s^* = [w^*, h^*] = \size(C^*)$.
    By the definition of $\sigma$, we have $\bar w \leq \hat w$ and $w^* \leq \hat w$.
    On the other hand, since $B_1(x_1,y_1) \subseteq \bar C$, we have $\bar w \geq w_1 = \hat w$. 
    By the construction of $C^*$, we have $w^* \geq w_1 = \hat w$.
    Summarizing the above, we have $\bar w = w^* = w_1$.
    
    By the assumption that $\bar C \in \Sset \setminus \argmin_{C \in \Sset} \{\size(C)\}$, we must have $\bar h > h^*$.    
    By the construction of $C^*$, we have $h^* \geq h_1$.
    Since $\bar C$, $C^*$ and $B_1$ are all regular, $\bar h$, $h^*$, and $h_1$ are all powers of $q_2$, hence $h^* | \bar h$ and $h_1 | h^*$. 
    Let $(\bar x, \bar y) = \location(\bar C)$.
    Consider the sets of regions $\Rset = \{\region(\bar x, \bar y + h^* \cdot j, w^*, h^*): j = 0,1,\ldots, \frac{\bar h}{h^*} - 1\}$. By construction, each $R \in \Rset$ is regular, aligned, and non-overlapping, and $\bigcup_{R \in \Rset} R = \bar C$. Furthermore, for each $R \in \Rset$, we have $\size(R) = \size(C^*) \succeq \size(B_1)$.
    By~\cref{lemma:overlap}, for each $R \in \Rset$, either $R \cap B_1(x_1,y_1) = \emptyset$ or $R \cap B_1(x_1,y_1) = B_1(x_1,y_1)$. However, $B_1(x_1,y_1) \subseteq \bigcup_{R \in \Rset} R$. By the pigeonhole principle, there must exists $R^* \in \Rset$ for which $B_1(x_1,y_1) \subseteq R^*$. 
    
    We show that a block is either contained in $R^*$ or it does not overlap with $R^*$ at all. More precisely, for all $j = 1, \ldots, m$, either $B_j(x_j,y_j) \cap R^* = \emptyset$ or $B_j(x_j,y_j) \cap R^* = B_j(x_j,y_j)$.
    
    Note that $w_j \leq \hat w = w^*$. 
    Suppose $h_j \leq h^*$, then we can use~\cref{lemma:overlap} 
    and conclude that either $B_j(x_j,y_j) \cap R^* = \emptyset$ or $B_j(x_j,y_j) \cap R^* = B_j(x_j,y_j)$.

    We now tackle the difficult case where $h_j > h^*$. 
    We first establish some notations.
    Recall that both $B_j$ and $R^*$ are regular, therefore both $h_j$ and $h^*$ are powers of $q_2$. Since $h_j > h^*$, we have $h^* | h_j$.
    Also recall that both $B_j(x_j,y_j)$ and $R^*$ are aligned. Let $(x^*,y^*) = \location(R^*)$. We have $h_j | y_j$ and $h^* | y^*$.
    Let $\alpha_j, \kappa_j, \kappa^*$ be non-negative integers such that $y^* = \kappa^* h^*$, $y_j = \kappa_j h_j = \alpha_j \kappa_j h^*$.
    Suppose towards contradiction that the lemma does not hold, then $[y^*, y^* + h^*) \cap [y_j, y_j + h_j) \neq \emptyset$.
    We argue that we must have $y_j \leq y^* < y^* + h^* \leq y_j + h_j$.
    Suppose this is the case, then $B_j(x_j,y_j) \cap B_1(x_1,y_1) \neq \emptyset$, contradicting the assumption that $\{(x_i,y_i)\}_{i=1}^{|\Bset|}$ is a solution to $(\Bset, \sigma(\Cset, \hat s))$, and hence the lemma is proven.
    
    Suppose the above assumption does not hold, then there are two possible cases. For the first case, we have $y^* < y_j <  y^* + h^* \leq y_j + h_j$.
    From the first two inequalities, we have $\kappa^* h^* < \alpha_j \kappa_j h^* < (\kappa^*+1) h^*$, which is impossible as there is no space to put an integer $\alpha_j \kappa_j$ between the consecutive integers $\kappa^*$ and $\kappa^* + 1$.
    Similarly, for the second case, we have $y_j \leq y^* < y_j + h_j < y^* + h^*$.
    From the last two inequalities, we have 
    $\kappa^* h^* < \alpha_j (\kappa_j+1) h^* < (\kappa^*+1) h^*$, which is again impossible.

    To recap, we have shown that for all $j$ we have either $B_j(x_j,y_j)$ is completely contained in $R^*$, or it does not overlap with $R^*$ at all. 
    This means that cutting the container $R^*$ from $\bar C$ does not invalidate the solution.
    Formally, we have that $(\Bset, (\sigma(\Cset,\hat s) \setminus \{\bar C\} )\cup \{\bar C \setminus R^*, R^* \})$ has a solution.
    Note that $R^*, C^* \in (\sigma(\Cset,\hat s) \setminus \{\bar C\} )\cup \{\bar C \setminus R^*, R^* \}$ and $\size(R^*) = \size(C^*)$.
    Therefore by~\cref{lem:swap} there exists a solution $\{(x^*_i,y^*_i)\}_{i=1}^{|\Bset|}$ to $(\Bset, (\sigma(\Cset,\hat s) \setminus \{\bar C\} )\cup \{\bar C \setminus R^*, R^* \})$ with $B_1(x^*_1,y^*_1) \subseteq C^*$. However, a solution to $(\Bset, (\sigma(\Cset,\hat s) \setminus \{\bar C\} )\cup \{\bar C \setminus R^*, R^* \})$ is also a solution to $(\Bset, \sigma(\Cset,\hat s))$ and therefore our claim is proven.
    \end{IEEEproof}

    \begin{claim}\label{claim:claim2}
        If there exists a solution $\{(x^*_i,y^*_i)\}_{i=1}^{|\Bset|}$ to $(\Bset, \sigma(\Cset,\hat s))$ with $B_1(x^*_1,y^*_1) \subseteq C^*$, then there also exists a solution $\{(x^\dagger_i,y^\dagger_i)\}_{i=1}^{|\Bset|}$ to $(\Bset, \sigma(\Cset,\hat s))$ with $(x^\dagger_1, y^\dagger_1) = \location(C^*)$.
    \end{claim}

\begin{IEEEproof}
    We will continue to assume that $w_1 = \hat w$ which implies that $w_1 = w^*$. As before, the case where $h_1 = \hat h$ can be dealt with using similar arguments. Now since $w_1 = w^*$, it must hold that $x^*_1 = x^\dagger_1$. Suppose $y^*_1 = y^\dagger_1$, then the claim is proven.
    Otherwise, suppose $y^*_1 \neq y^\dagger_1$. Note that $y^\dagger_1 < y^*_1 < y^*_1 + h_1 \leq y^\dagger_1 + h^*$. Let $\delta := y^*_1 - y^\dagger_1$, and let $(\delta_0, \ldots, \delta_\ell)$ be the $q_2$-ary representation of $\delta$ where $\delta = \sum_{j = 0}^{\ell} \delta_j q_2^{j}$ and $\ell = \log_{q_2} h^* - 1$.
    Let $j^* \leq \ell$ be the largest integer for which $\delta_{j^*} \neq 0$.
    We define the regions $R_{j^*} := \region(x^\dagger_1, y^\dagger_1, w^*, q_2^{j^*})$ and $R'_{j^*}:= \region(x^\dagger_1, y^\dagger_1 + \delta_{j^*} \cdot q_2^{j^*}, w^*, q_2^{j^*})$. It is straightforward to check that both $R_{j^*}$ and $R'_{j^*}$ are regular aligned. 
    
    Recall that $B_1$ and $C^*$ are regular, therefore $h_1$ and $h^*$ are both powers of $q_2$. Since $h_1 \leq h^*$, we have $h_1 | h^*$. Since $B_1(x^*_1,y^*_1)$ and $C^*$ are aligned, both $y^*_1$ and $y^\dagger_1$ are multiples of $h_1$. Therefore $\delta$ is also a multiple of $h_1$. Since we have assumed $\delta > 0$, it must be the case that $h_1 \leq \delta$. Since $h_1$ is a power of $q_2$, we have $h_1 \leq q_2^{j^*}$.
    Therefore, $B_1(x^*_1, y^*_1) \subseteq R'_{j^*}$.  
    
    Next, we argue that each block $B_j(x_j^*, y_j^*)$ is either contained in $R_{j^*}$, or contained in $R'_{j^*}$, or does not overlap with $R_{j^*}$ and $R'_{j^*}$ at all. We first note that $\size(R_{j^*}) = \size(R'_{j^*}) = [w^*, q_2^{j^*}]$, and $w_j \leq w_1 = w^*$. Therefore if $h_j \leq q_2^{j^*}$ we can use~\cref{lemma:overlap} and prove the claim. 
    
    Now consider the case where $h_j > q_2^{j^*}$. Since $B_j$ is regular, we have $h_j \geq q_2^{j^* + 1}$. Note that the lower boundary of $R_{j^*}$ is located at $y^\dagger_1$ unit of the second dimension, while the lower boundary of $R'_{j^*}$ is located at $y^\dagger_1 + \delta_{j^*} \cdot q_2^{j^*}$ unit. The two boundaries therefore have a height difference of $\delta_{j^*} \cdot q_2^{j^*} < q_2^{j^*+1}$ unit. Thus $y_j$ cannot be in between the two boundaries or else $B_j(x_j^*, y_j^*)$ would overlap with $B_1(x_1^*, y_1^*)$. In other words, we have either $y^*_j < y^\dagger_1$ or $y^*_j \geq y^\dagger_1 + \delta_{j^*} \cdot q_2^{j^*}$.
    
    Consider the first case where $y^*_j < y^\dagger_1$. If $y^\dagger_1 < y^*_j + h_j$, then part of $B_j$ is contained by $C^*$ and part of it is contained by some other container(s), which violates the assumption that $\{(x^*_i,y^*_i)\}_{i=1}^{|\Bset|}$ is a solution to $(\Bset, \sigma(\Cset,\hat s))$. We must therefore have  $y^*_j + h_j \leq y^\dagger_1$. This implies that $B_j(x^*_j, y^*_j)$ does not overlap with $C^*$. Hence it also does not overlap with $R_{j^*}$ and $R'_{j^*}$.
    
    Now consider the second case where $y^*_j \geq y^\dagger_1 + \delta_{j^*} \cdot q_2^{j^*}$. Suppose that $y^\dagger_1 + \delta_{j^*} \cdot q_2^{j^*} \leq y^*_j < y^\dagger_1 + (\delta_{j^*} + 1) \cdot q_2^{j^*}$. Recall that $R'_{j^*}$ and $B_j$ are both regular, therefore $y^\dagger_1 = \mu q_2^{j^*}$ and $y^*_j = \nu q_2^{j^*}$ (since $h_j$ is a multiple of $q_2^{j^* + 1}$) for some non-negative integers $\mu$ and $\nu$. Substituting the expressions, we have  
    \begin{align*}
    y^\dagger_1 + \delta_{j^*} \cdot q_2^{j^*} \leq y^*_j < y^\dagger_1 + (\delta_{j^*} + 1) \cdot q_2^{j^*} \\
    (\mu + \delta_{j^*}) q_2^{j^*} \leq \nu q_2^{j^*} < (\mu + \delta_{j^*} + 1) q_2^{j^*}
    \end{align*}
    which forces $\nu = \mu + \delta_{j^*}$ and hence $y^\dagger_1 + \delta_{j^*} \cdot q_2^{j^*} = y^*_j$. This however contradicts to the assumption that $\{(x^*_i,y^*_i)\}_{i=1}^{|\Bset|}$ is a solution to $(\Bset, \sigma(\Cset,\hat s))$, since $B_j(x^*_j,y^*_j)$ overlaps with $B_1(x^*_1,y^*_1)$. Therefore it must be the case that $y^*_j \geq y^\dagger_1 + (\delta_{j^*} + 1) \cdot q_2^{j^*}$ and therefore $B_j(x^*_j, y^*_j)$ does not overlap with $R_{j^*}$ and $R'_{j^*}$.
    
    To recap, we have constructed two regular aligned regions $R_{j^*}$ and $R'_{j^*}$ of the same size, such that they are contained in $C^*$ and $R'_{j^*}$ contains $B_1(x^*_1, y^*_1)$. Furthermore, each block $B_j(x^*_j,y^*_j)$ is either contained in  $R_{j^*}$, or in $R'_{j^*}$, or does not overlap with them at all. From these guarantees, we can conclude that $\{(x^*_i,y^*_i)\}_{i=1}^{|\Bset|}$ is also a solution to $(\Bset, (\sigma(\Cset, \hat s) \setminus \{C^*\}) \cup \{R_{j^*}, R'_{j^*}, C^* \setminus (R_{j^*} \cup R'_{j^*})  \} )$. We can then apply~\cref{lem:swap} and obtain a solution to $(\Bset, (\sigma(\Cset, \hat s) \setminus \{C^*\}) \cup \{R_{j^*}, R'_{j^*}, C^* \setminus (R_{j^*} \cup R'_{j^*})  \} )$, which is also a solution to $(\Bset, \sigma(\Cset, \hat s))$, such that $B_1$ is packed into $R_{j^*}$ instead.
    Repeating this process until there does not exist $j^*$ with $\delta_{j^*} \neq 0$, we can obtain a solution in which $(x^\dagger_1, y^\dagger_1)$ is assigned to $B_1$.
\end{IEEEproof}

Lemma~\ref{lem:core} states a way to formulate the problem after we packed $B_1$ by packing $B_1$ at the bottom left corner of $C^\ast$, where $C^\ast$ is a smallest container at hand which is large enough to contain $B_1$.\footnote{To see why, we give a simple example.
Suppose we have $q_1 = q_2 = 2$, two blocks of sizes $2 \times 1$ and $1 \times 2$, and two aligned containers of sizes $2 \times 2$ and $2 \times 1$.
It is easy to see that the problem has a solution.
If we choose the $2 \times 2$ container, which is not the smallest, and pack the largest block, \ie, the $2 \times 1$ block, into it, then what we left are two $2 \times 1$ containers.
There is no way to pack the remaining $1 \times 2$ block.}
We present Fig.~\ref{fig:lowest} and \ref{fig:smallest} to give a high level explanation on why Lemma~\ref{lem:core} works.
In these figures, we let $q_1 = q_2 = 2$, and the red block be the largest block at hand.
After applying Lemma~\ref{lem:cut}, the red block must fit to either the width or the height of the container.
Here we illustrates the case that the width is fitted.
Suppose $(\Bset, \Cset)$ has a solution, then $(\Bset, \sigma(\Cset, s^\ast))$ also has a solution by Lemma~\ref{lem:cut}.
Let $\bar{C}$ be the container $B_1$ is packed in a solution to $(\Bset, \sigma(\Cset, s^\ast))$.
If $\bar{C} \in \Sset$, then without loss of generality we can say $C^\ast = \bar{C}$.\footnote{It is trivial that if $C^\ast$ and $\bar{C}$ are distinct containers in $\Sset$, we can swap the blocks contained by them without affecting the validity of the solution as they are both regular aligned containers of the same size.}
We can swap the blocks in $C^\ast$ such that the largest block can be packed at the bottom left corner while preserving the validity of the solution, which is shown in Fig.~\ref{fig:lowest}.
If $\bar{C} \not\in \Sset$, then we have $\size(\bar{C}) \succeq \size(C^\ast)$.
Note that one of the side of $\bar{C}$ has the same length as the one of $C^\ast$.
Fig.~\ref{fig:smallest} illustrates that we can swap the blocks in $C^\ast$ and some of the blocks in $\bar{C}$ while preserving the validity of the solution.

\subsection{Correctness of Our Algorithm and Time Complexity}

With the help of the established lemmas, we are now able to prove Theorem~\ref{thm:main}.

\begin{IEEEproof}(Theorem~\ref{thm:main})
It is straightforward to check that if Algorithm~{alg:csp} terminates without outputting $\bot$, then its output is a valid solution to the problem $(\Bset, \Cset)$.
The converse of this theorem is the main result of this paper, which states that if $(\Bset, \Cset)$ indeed has a solution, then Algorithm~{alg:csp} can find one of the solutions.
We prove by recursively applying~\cref{lem:core}.

By~\cref{lem:core}, if $(\Bset, \Cset)$ has a solution, then there must exist a solution in which the first (which is the largest) block is packed into certain containers. Specifically, using the notations in~\cref{lem:core}, for any container $C^*$ which is among the smallest containers in $\sigma(\Cset, \hat s)$ which are large enough to contain $B_1$, there must exist a solution $\{(x^*_i,y^*_i)\}_{i = 1}^{|\Bset|}$ where $(x^*_1, y^*_1) = \location(C^*)$. Note that the first iteration of Algorithm~{alg:csp} picks exactly one such $C^*$ and sets $(x^*_1, y^*_1) = \location(C^*)$. Using again the notations in~\cref{lem:core}, we let $\Bset'$ be the multiset of blocks obtained from $\Bset$ by removing $B_1$, and $\Cset'$ be the set of containers obtained from $\Cset$ by removing the space occupied by $B_1$. Note that the second largest block in $\Bset$ is now the largest block in $\Bset'$. We can then apply~\cref{lem:core} again on $(\Bset',\Cset')$ which guarantees that Algorithm~{alg:csp} assigns a good location for $B_2$. Repeating this process, Algorithm~{alg:csp} is able to assign a good location to each block in $\Bset$, and the lemma is thus proven. 
\end{IEEEproof}

Given a set $\Cset$ of containers and
a set $\Bset = \{B_1, \ldots B_m\}$ of blocks.
Define $[w_B^i, h_B^i] := \size(B_i)$ for $i = 1, \ldots, m$, $w_\text{max} := \max_{i = 1}^m w_B^i$ and $h_\text{max} := \max_{i = 1}^m h_B^i$.
Observe that we 
only need to keep track of the number of containers of different sizes, we create a 2D array $\mathbf{A}$, where $\mathbf{A}_{i,j}$ stores the number of containers of size $[q_1^i, q_2^j]$ for $i \in \{0, \ldots, \log_{q_1} w_\text{max}\}$, $j \in \{0, \ldots, \log_{q_2} h_\text{max}\}$.
This array can be initialized in $\mathcal{O}((\log_{q_1} w_\text{max})(\log_{q_2} h_\text{max}))$ time.
On the other hand, for each input container of size $[a,b]$, 
we can perform the initial cut by $\sigma$ and register the cut containers to $\mathbf{A}$ in $\mathcal{O}((\log_{q_1} a)(\log_{q_2} b))$ time.

\begin{definition}[Layers]
	A block or a container of size $[w,h]$ belongs to \emph{layer} $l$ if and only if $l = \max\{w,h\}$.
\end{definition}

In Algorithm~\ref{alg:csp}, the blocks are packed sequentially according to their sizes sorted in descending order.
That is, the algorithm packs the blocks belonging to the same layer before it packs the blocks belonging to a lower layer.
When the algorithm handles the layer $l$, every container $C \in \Cset$ has $\size(C) \preceq [l,l]$.
We can ensure that every container which can contain some block belonging to layer $l$ has at least one of its sides fitting the corresponding side of the block.
That is, when the algorithm finishes packing all the blocks belonging to layer $l$ and moves to layer $l'$, we only need to apply $\sigma(\Cset,[ q_1^{\lfloor \log_{q_1} l' \rfloor}, q_2^{\lfloor \log_{q_2} l' \rfloor} ])$ instead of the one shown in the algorithm.
The actual procedure is to move the count from $\mathbf{A}_{i,j}$ to either $\mathbf{A}_{i-k,j}$, $\mathbf{A}_{i,j-p}$ or $\mathbf{A}_{i-k,j-p}$ for some $k \in \{1,2,\ldots,i\}, p \in \{1,2,\ldots,j\}$, where the corresponding $i, j$ representing the layer $l = \max\{i,j\}$.
We can actually consider $k = p =1$, as if no block belongs to the new layer, we can simply move again to a lower layer.
For each case when $k = p = 1$, we only need to multiply $\mathbf{A}_{i,j}$ by $q_1$, $q_2$ or $q_1 q_2$ respectively and add it to the count stored in the corresponding new position, which takes $\mathcal{O}(1)$ time.
That is, the total time complexity of $\sigma$ for moving from a layer to a lower one during the whole algorithm is $\mathcal{O}((\log_{q_1} w_\text{max})(\log_{q_2} h_\text{max}))$.

We now consider the block assignment in layer $l$.
Without loss of generality, we assume that the largest block at hand has size $[l, m]$ for some $m \le l$.
The only containers which can contain the block must be in the same layer.
If there is a container with size equals to the size of the block, \ie, $\mathbf{A}_{\log_{q_1} l, \log_{q_2} m} > 0$, then that container is accountable for the block assignment since it is the smallest container which can contain the block.
For the case $\mathbf{A}_{\log_{q_1} l, \log_{q_2} m} = 0$, we look for the smallest $j > \log_{q_2} m$ such that $\mathbf{A}_{\log_{q_1} l, j} > 0$, which takes $\mathcal{O}(\log_{q_2} l)$ time by sequential search.
Recall that the blocks are sorted according to their size in descending order.
Due to the fact that an $l \times q_2^j$ container can be used to pack $q_2^{j}/m$ number of $l \times m$ blocks, we can perform successive assignments of $l \times q_2^j$ blocks without actually applying $\sigma$ to the container after each assignment.
After all the $l \times m$ blocks are packed, if the next block to be packed has size $l \times (m/q_2^k)$ for some $k$, then the successive assignment for the new block still works.
Hence, we can simulate $\sigma$ by simple subtraction until the width of the new block is no longer $l$.
A similar argument can be applied when the largest block at hand has size $[m, l]$. 

We now calculate an upper bound on the time complexity of Algorithm~\ref{alg:csp}.
We have to search a container for every block, where the worst case for a  search can be upper bounded by $\mathcal{O}(\max\{\log_{q_1} w_\text{max}, \log_{q_2} h_\text{max}\})$.
At the end of a successive assignment sequence, we have to cut the unused region of the container, where the cut takes either $\mathcal{O}(\log_{q_1} l)$ or $\mathcal{O}(\log_{q_2} l)$ time depending on whatever we cut the width or the height.
The time complexity for the cuts after successive assignments is upper bounded by $\mathcal{O}(\log_{q_1} w_\text{max} + \log_{q_1} (w_\text{max}/q_1) + \ldots + \log_{q_1} q_1 + \log_{q_2} h_\text{max} + \log_{q_2} (h_\text{max}/q_2) + \ldots + \log_{q_2} q_2) = \mathcal{O}(\log_{q_1}^2 w_\text{max} + \log_{q_2}^2 h_\text{max})$.
The time of $\sigma$ for moving the layer takes $\mathcal{O}((\log_{q_1} w_\text{max})(\log_{q_2} h_\text{max}))$, which can be absorbed by $\mathcal{O}(\log_{q_1}^2 w_\text{max} + \log_{q_2}^2 h_\text{max})$.
Let $W_\text{max}$ and $H_\text{max}$ be the maximum width and height among the input containers.
The overall time complexity is $\mathcal{O}(m \max\{\log_{q_1} w_\text{max}, \log_{q_2} h_\text{max}\} + \log_{q_1}^2 w_\text{max} + \log_{q_2}^2 h_\text{max} + |\Cset| (\log_{q_1} W_\text{max})(\log_{q_2} H_\text{max}))$.

At last, we show the time complexity when we apply the rectangle packing problem to solve our original goal: Decide whatever a two-channel prefix code exists for a given multiset of codeword lengths.
Note that $\ell_1^\text{max} = \log_{q_1} w_\text{max}$ and $\ell_2^\text{max} = \log_{q_2} h_\text{max}$.
At the beginning, we only have one regular aligned container $\region(0, 0, w_\text{max}, h_\text{max})$.
Including the time for sorting, the time complexity is $\mathcal{O}(m \log m + m \max\{\ell_1^\text{max}, \ell_2^\text{max}\} + (\ell_1^\text{max})^2 + (\ell_2^\text{max})^2)$.

\section{Concluding Remarks}

In this paper, we closed the gap where the two-channel Kraft inequality fails by formulating a decision procedure for the existence of two-channel prefix codes via solving a rectangle packing problem with dimension-wise alignment constraints.

\appendix

\section{Proofs of Theorem~\ref{thm:kraft} and Theorem~\ref{thm:entropy}} \label{sec:kraft_entropy}

We first prove the Kraft inequality.

\begin{IEEEproof}
	Let $N$ be an arbitrary positive integer.
	Consider
	\begin{IEEEeqnarray*}{Cl}
		& \left( \sum_{j = 1}^m \prod_{i = 1}^n q_i^{-\ell_i^j} \right)^N = \sum_{j_1 = 1}^m \sum_{j_2 = 1}^m \ldots \sum_{j_N = 1}^m \left( \prod_{i = 1}^n q_i^{-\sum_{k = 1}^N \ell_i^{j_k}} \right) \\
		= & \sum_{k_1 = 1}^{N\ellmax_1} \sum_{k_2 = 1}^{N\ell_2^\text{max}} \ldots \sum_{k_n = 1}^{N\ellmax_n} A_{k_1, k_2, \ldots, k_n} \prod_{i = 1}^n q_i^{-k_i}, 
        \label{eq:kraft_expand}
	\end{IEEEeqnarray*}
	where $A_{k_1, k_2, \ldots, k_n}$ is the coefficient of $\prod_{i = 1}^n q_i^{-k_i}$ in $( \sum_{j = 1}^m \prod_{i = 1}^n q_i^{-\ell_i^j} )^N$.

	Note that $A_{k_1, k_2, \ldots, k_n}$ gives the total number of sequences of $N$ codewords with a total length of $k_i$ symbols in the $i$-th channel for all $i = 1, 2, \ldots, n$.
	Since the code is uniquely decodable, these code sequences must be distinct.
	So, the number $A_{k_1, k_2, \ldots, k_n}$ must be no more than the total number of distinct sequences where there are $k_i$ symbols in the $i$-th channel for all $i = 1, 2, \ldots, n$.
	That is,
	\begin{equation} \label{eq:kraft_count}
		A_{k_1, k_2, \ldots, k_n} \le \prod_{i = 1}^n q_i^{k_i}.
	\end{equation}

	Now, substitute \eqref{eq:kraft_count} into \eqref{eq:kraft_expand} and get
	\begin{equation*}
		\left( \sum_{j = 1}^m \prod_{i = 1}^n q_i^{-\ell_i^j} \right)^N \le \sum_{k_1 = 1}^{N\ellmax_1} \sum_{k_2 = 1}^{N\ell_2^\text{max}} \ldots \sum_{k_n = 1}^{N\ellmax_n} 1.
	\end{equation*}

	Since this inequality holds for any $N$, so we let $N \to \infty$ and obtain our desired Kraft inequality.
\end{IEEEproof}

With the help of the Kraft inequality, we can now prove the entropy bound.

\begin{IEEEproof}
	Recall that $H_D(Z) = -\sum_{j = 1}^m p_j \log_D p_j$.
	Consider
	\begin{IEEEeqnarray*}{Cl}
		& \sum_{j = 1}^m p_j \sum_{i = 1}^n \log_D q_i^{\ell_i^j} - H_D(Z)\\
		= & \sum_{j = 1}^m p_j \log_D \left( \prod_{i = 1}^n q_i^{\ell_i^j} \right) + \sum_{j = 1}^m p_j \log_D p_j\\
		= & (\ln D)^{-1} \sum_{j = 1}^m p_j \ln \left( p_j \prod_{i = 1}^n q_i^{\ell_i^j} \right)\\
		\ge & (\ln D)^{-1} \sum_{j = 1}^m p_j \left( 1 - \left( p_j \prod_{i = 1}^n q_i^{\ell_i^j} \right)^{-1} \right) 
        \label{eq:entropy_ln} \\
		= & (\ln D)^{-1} \left( 1 - \sum_{j = 1}^m \prod_{i = 1}^n q_i^{-\ell_i^j} \right)\\
		\ge & 0, 
        \label{eq:entropy_kraft}
	\end{IEEEeqnarray*}
	where \eqref{eq:entropy_ln} follows the inequality $\ln a \ge 1 - 1/a$ for any $a > 0$, and \eqref{eq:entropy_kraft} follows the Kraft inequality \eqref{eq:kraft}.

	The equality in \eqref{eq:entropy_ln} holds if and only if $p_j \prod_{i = 1}^n q_i^{\ell_i^j} = 1$ for all $j$, or equivalently, $\sum_{i = 1}^n \log_D q_i^{\ell_i^j} = -\log_D p_j$ for all $j$.
	Under this condition, we have
	\begin{equation*}
		\sum_{j = 1}^m \prod_{i = 1}^n q_i^{-\ell_i^j} = \sum_{j = 1}^m p_j = 1.
	\end{equation*}
	So, the equality in \eqref{eq:entropy_kraft} also holds, which means that the bound is tight.
\end{IEEEproof}

\bibliographystyle{IEEEtran}
\bibliography{packing-bib}

\end{document}